\documentclass{tcibook}
\usepackage{fancyhea}
\usepackage{work}
\usepackage{bm}       
\usepackage{graphicx}
\usepackage{multirow}
\usepackage{lineno}
\usepackage{threeparttable}
\usepackage[normalem]{ulem}
\usepackage{array}
\usepackage{deluxetable}
\usepackage{color}
\usepackage[colorlinks = true,
            linkcolor = blue,
            urlcolor  = blue,
            citecolor = blue,
            anchorcolor = blue]{hyperref}
\usepackage{amsmath,amssymb}
\usepackage[titles]{tocloft}
\def\ie{{\it i.e.}}
\def\eg{{\it e.g.}}

\def\beq{\begin{equation}}
\def\eeq#1{\label{#1}\end{equation}}
\def\eeqn{\end{equation}}

\newenvironment{Eqnarray}%
   {\arraycolsep 0.14em\begin{eqnarray}}{\end{eqnarray}}
\def\beqa{\begin{Eqnarray}}
\def\eeqa#1{\label{#1}\end{Eqnarray}}
\def\eeqan{\end{Eqnarray}}

\let\bar=\overbar

\def\lsim{\mathrel{\raise.3ex\hbox{$<$\kern-.75em\lower1ex\hbox{$\sim$}}}}
\def\gsim{\mathrel{\raise.3ex\hbox{$>$\kern-.75em\lower1ex\hbox{$\sim$}}}}

\def\del{\partial}
\def\Dslash{\not{\hbox{\kern-4pt $D$}}}
\def\dslash{\not{\hbox{\kern-2pt $\del$}}}
\def\pslash{\not{\hbox{\kern-2pt $p$}}}
\def\ETmiss{\not{\hbox{\kern-4pt $E$}}_T}

\def\Dlr{\mathrel{\raise1.5ex\hbox{$\leftrightarrow$\kern-1em\lower1.5ex\hbox{$D$}}}}

\def\MSB{{\bar{M \kern -2pt S}}}
\def\msb{{\bar{\scriptsize M \kern -1pt S}}}

\def\drb{{\bar{\scriptsize D \kern -1pt R}}}

\begin{document}


\pagenumbering{roman}

\parindent=0pt
\parskip=8pt
\setlength{\evensidemargin}{0pt}
\setlength{\oddsidemargin}{0pt}
\setlength{\marginparsep}{0.0in}
\setlength{\marginparwidth}{0.0in}
\marginparpush=0pt


\pagenumbering{arabic}

\renewcommand{\chapname}{chap:intro_}
\renewcommand{\chapterdir}{.}
\renewcommand{\arraystretch}{1.25}
\addtolength{\arraycolsep}{-3pt}

\setcounter{chapter}{3} 
\setcounter{tocdepth}{2}

\newcommand{\np}{$\bar{n} P(k)$}
\newcommand{\chimp}{$h^{-1}$\,Mpc (comoving)}
\newcommand{\lcdm}{$\Lambda$CDM}
\newcommand{\neff}{$N_{\rm eff}$} 
\newcommand{\fnlloc}{$f_{\rm nl}^{\rm loc}$}
\newcommand{\fnl}{$f_{\rm nl}$}

\newcommand{\as}[1]{{\textcolor{red}{[\textbf{AS}: #1]}}}
\newcommand{\ja}[1]{{\textcolor{green}{[\textbf{JA}: #1]}}}
\newcommand{\jn}[1]{{\textcolor{blue}{[\textbf{JN}: #1]}}}


\chapter{Dark Energy and Cosmic Acceleration in the Modern Universe}

\vskip 0.2in

\begin{center}
{\bf \Large Conveners: James Annis, Jeffrey A. Newman, and An\v{z}e Slosar} 
\\
\vspace*{0.2cm}
{\large Contributors: Jonathan Blazek, Sukanya Chakrabarti, Kyle Dawson, Alex Drlica-Wagner, Steve Eikenberry, Simone Ferraro,  Anthony H. Gonzalez,  Andrew Hearin, Katrin Heitmann, Alex Kim, Rachel Mandelbaum, Jeffrey McMahon, Jessica Muir, Peter Nugent, Aaron Roodman, Noah Sailer, Benjamin Wallisch, Martin White, and Rongpu Zhou
}

\end{center}


\let\clearpage\relax
\tableofcontents

\section{Executive Summary}

In the 25 years since the discovery of dark energy  new technologies, theoretical advances, and powerful
experiments have produced a tremendous advance in our understanding of cosmology. However, fundamental
questions remain open. What is the nature of dark energy? Is general relativity correct at all scales and at
all times? What is the nature of dark matter and how does it connect to the standard model of particle
physics? Can we show how inflation established the initial conditions for the Universe as we observe it today?
If so, what can we learn about the fields driving inflation and how do these connect to the standard model of particle physics? This topical group report describes the science our community believes will best achieve progress in our understanding of these questions over the next decades using observations of the $z<6$ universe.

Data from the modern universe has played a key role in our attempts to answer these fundamental questions about the nature and origins of the observable universe. There is a tremendous opportunity to make progress on these questions by pursuing large galaxy redshift surveys spanning the era following the epoch of reionization ($z \lesssim 6$). 
The techniques of precision cosmology pioneered in experiments over the last decades at $z \lesssim 1$ can be applied to galaxies out to cosmic dawn, providing access to a factor of 10 more volume during a time when the universe was simpler.

The community's input to this report jointly describes a multi-probe experimental program that provides a rich, deep, and flexible portfolio that provides powerful constraints on cosmic acceleration. The program
includes a powerful new Stage V spectroscopic facility (also referred to here as Spec-S5) that would pursue
larger and deeper surveys enabling transformational advances in our understanding of \textit{both} eras of
accelerated expansion in the history of the universe -- the early inflationary epoch and the late dark energy-driven one.   With such data, we can put tight constraints on the behavior of dark energy out to the redshift where it ceases to be dynamically important in the \lcdm\ model; 
we can test the statistics of the primordial fluctuations sourcing density variations and thus constrain many inflationary models;
and we can look for features in the primordial matter power spectrum that would provide new information on the physics of inflation. 
The same datasets would simultaneously enable us to place stringent constraints on dark radiation, ruling out entire classes of dark sector models, and allow us to  measure the sum of the neutrino masses from their effect on the growth of overdensities.  Some of the opportunities provide natural target levels of precision for enabling new discoveries, while others -- such as whether the evidence for the accelerated expansion of the universe is really an indication that general relativity fails at large scales and low accelerations -- are part of a broad program to make progress on difficult questions.

A Stage V spectroscopic facility would be a flexible science machine for many areas of modern cosmology and astrophysics. As discussed elsewhere in this report, it would be particularly powerful for constraining the nature of dark matter via astrophysical and cosmological probes. 

 
Our community identified two key programs for a Stage V spectroscopic facility studying acceleration:
\begin{itemize}
\item \textbf{Lower-redshift ($\mathbf{z<1.5}$), high-density spectroscopic surveys tracing non-linear scales: }\\
Dramatically increasing the number of galaxies with  measurements at lower redshift will provide datasets that are highly sensitive to the properties of dark matter and to minute departures from the general relativity while providing multiple cross-checks that will make results robust to systematic effects. Additionally, this data will measure large-scale modes with precision far exceeding what is required for basic analyses and thus enable precision multi-tracer studies. 


\item \textbf{High-redshift ($\mathbf{z\gtrsim 2}$), high-volume spectroscopic surveys tracing linear scales:}\\
A next-generation high-redshift survey could sample large volumes beyond redshift of $z>2$ to maximize the number of well-measured \emph{linear} modes.  This will enable extremely sensitive measurements of early universe physics, including providing tests for primordial non-Gaussianity and features in the matter power spectrum as well as constraints on early dark-energy models. 
\end{itemize}
\vspace{-1em}
Acquiring these samples will require implementation of a highly-multiplexed spectrograph on a new,
large-aperture ($\gtrsim$ 6\,m), wide-field-of-view telescope located appropriately for targeting sources in
the deepest large area photometric survey in the coming decade, the Vera C. Rubin Observatory Legacy Survey of
Space and Time (LSST).  Proposals for such a facility include the Maunakea Spectroscopic Explorer, MegaMapper,
and European Southern Observatory SpecTel concepts. Until the availability of such a facility, DESI will
provide a unique resource for large, wide-field surveys. As a result, the Dark Energy Spectroscopic Instrument (DESI) should play a key role in prototyping Stage V surveys and obtaining data to complement LSST through the late 2020s at minimum, producing spectroscopic samples that address the most compelling science opportunities in a more limited way than Spec-S5 can.

In addition to the science opportunities enabled by a new spectroscopic facility, our community has identified other promising areas for future work that can efficiently employ current-generation facilities or be undertaken as smaller programs:


\begin{itemize}

\item \textbf{Supporting science analyses from Stage IV facilities:}\\ The coming decade will present great opportunities to make progress on the problem of cosmic acceleration thanks to the portfolio of Stage IV dark energy experiments coming online. To take maximum advantage of the DESI and LSST datasets the science collaborations analyzing these data will need funding for both scientific infrastructure and research activities.  Cross-survey simulations and combined analyses of multiple experiments provide opportunities to extract maximal cosmological information, but will require additional support and/or new modes of working to succeed. 

\item \textbf{Enhancing the science gains from near-future facilities via additional data:}\\
The science return from upcoming experiments, particularly the LSST, can be greatly enhanced with modest investment into follow-up observations, including small-aperture telescopes, large telescopes, and the upcoming generation of Extremely Large Telescopes. \emph{Photometric redshifts} will likely constitute the limiting systematic for LSST;  their performance can be enhanced via dedicated calibration surveys executed on existing instruments and/or a Stage V spectroscopic facility, greatly improving the cosmological constraining power of LSST data. \emph{Supernova cosmology} requires the use of follow-up facilities, both in a time-sensitive manner to accurately determine the classifications for a subset of LSST supernovae as well as non-time-sensitive redshifting of supernova host galaxies.  \emph{Strong lensing cosmology} requires follow-up imaging to measure quasar light-curves in LSST-identified systems, high-priority observations of any strongly-lensed SN candidates, and adaptive optics IFU spectroscopy to measure source positions and lens galaxy velocity dispersions. \emph{Peculiar velocities of low-redshift Supernovae} provide a novel tool for measuring the structure of the universe at the lowest redshifts, but requires assembling data from a network of modestly-sized telescopes to measure precision light-curves and measure spectra of those objects. \emph{Standard Siren cosmology} requires redshifts to make use of the first principles distances measured by LIGO, Virgo, and soon Kagara as well as from the next generation of gravitational wave observatories. 

\item \textbf{Future use of the Vera Rubin Observatory:} \\The program for Rubin Observatory after completion of LSST should be evaluated in a dedicated exercise later in this decade once the performance of LSST and the progress of other projects is known. Options include new surveys with a different focus (cadence, area, depth, etc.), modest instrumentation changes such as a new filter set, or more radical alterations to the camera and/or telescope system. 

\item \textbf{Research and development to enable future probes of cosmology:}\\
There are a number of areas where investment now could enable novel and potentially powerful cosmological probes to begin operations in future decades. \emph{High-precision optical spectroscopy} techniques are being developed that would enable direct measurements of the expansion rate of the universe in the next few decades. \emph{High-precision astrometry} methods are also now being developed that would enable novel dark matter and cosmological probes by observing the evolution of the universe in real time. Some of these techniques include novel uses of quantum measurement methods that are being explored and funded in other contexts. \emph{21\,cm Spectroscopy and millimeter-wave Line Intensity Mapping} are novel techniques that have the potential to transform our ability to measure large-scale structures in the universe by observing the aggregate fluctuations in intensity rather than individual objects, but require further investments to become competitive.
\end{itemize}\vspace{-0.5em}

As described in this report, the modern universe is rich in cosmological information, providing many opportunities to improve our understanding of fundamental physics, including exploring the nature of cosmic acceleration but extending far beyond.  

\section{Introduction: Key Physics Questions and Opportunities}

Our understanding of cosmology underwent a phase transition in the early 2000s, when the experimental data changed our knowledge of cosmological parameters from order of magnitudes estimates to $10\%$ level constraints. The 6-parameter \lcdm\ model has emerged as the standard cosmological model. \lcdm\  has proven to be generally consistent with a large compendium of experimental data. Measurements advanced rapidly: statistical uncertainties halved with every generation of experiments, while at the same time remaining  large enough that only the largest and most obvious systematic errors were important. 
Over the decade of the 2010s, the community coalesced around larger experiments and collaborations in order to make progress. New, more complex datasets required processing by sophisticated codes developed by teams of scientists.  Our ability to constrain cosmological models has continued to grow rapidly; today we know the value of some cosmological quantities, such as the age of the Universe, with percent-level precision. 

Still, many key open questions about our universe remain unresolved:
\begin{itemize}
\item \emph{What is dark energy?} By the early 2000s, it was clear that some unknown component must drive the accelerating expansion of the universe if general relativity is the correct description of gravity; today, we know that this component must have an equation of state that matches that of a vacuum energy at the few percent level. However, the space of theories that are consistent with observations remains large. 

\item \emph{What is the nature of cosmic inflation?} It is still unknown if inflation left observable footprints that will help us to discern details of its mechanism. Through stringent upper limits on the amplitude of the primordial gravitational waves, we know that the simplest inflationary models (consisting of scalar field with a quadratic potential) are ruled out. Future observations of inflationary relics in the density fields observed by  next-generation spectroscopic surveys could transform our understanding of the Universe. 

\item \emph{What is dark matter?} While the constraints on deviations from a cold, interaction-less fluid have improved by orders of magnitude, we have not yet found a smoking gun of the new physics responsible for the presence of dark matter, nor what fundamental physics governs its properties. 

\item \emph{Is gravity well-described by Einstein's general relativity, or do we need new degrees of freedom
    to describe its action on cosmic scales?} General relativity  can describe observed phenomena over 30
        orders of magnitude in distance scale, ranging from sub-millimeter force measurements in the
        laboratory to the largest observable scales in the Universe. Yet, theoretical investigations continue
        explore new avenues that may allow us to ultimately connect it to the standard model of particle physics.  Modifications to general relativity could offer an alternative explanation for the observed cosmic acceleration.
\end{itemize}

There are tantalizing tensions in the current set of observations that, if not due to systematics, may hint of new physics.
Measurements made at early and late times in the history of the Universe are in tension with each other. 
The present-day cosmic expansion rate inferred from observations of CMB combined with lower-$z$ probes is a few percent inferred from cosmological observations of the CMB or large-scale structure of the universe is a few percent lower than that inferred via the direct distance ladder; 
the statistical significance of this difference in some cases exceeds 5$\sigma$. Similarly, the amplitude of cosmic density fluctuations (as measured by the $\sigma_8$ parameter) inferred from weak lensing and other probes at lower $z$ is low compared to extrapolations from the high-$z$ universe. While the tension in these measurements is less strong, its statistical significance is getting uncomfortably high. 

These tensions did not appear in sectors where we expect them to appear and besides systematic errors in the
experirments they do not have a natural explanation; i.e., we have not yet found a convincing theoretical explanation.
They point towards new, promising, research directions that build on the techniques developed to study cosmic acceleration in the modern universe, which we describe in this chapter. If these tensions are real, they will provide a wonderful opportunity to learn something fundamentally new. 

\subsection{Unexpected Discoveries}

Many cosmological techniques proposed during the 2000s have now borne fruit. The two most prominent are weak gravitational lensing and baryon acoustic oscillations (BAO). These measurements have confirmed the standard model of cosmology and in particular the accelerated expansion of the Universe as discovered a decade earlier by supernova Type Ia measurements. Weak gravitational lensing was identified as the method of choice for constraining dark energy in the 2005 report of the Dark Energy Task Force \cite{Albrecht2006}, and in 2022 forms the core of the DES Y3 cosmology results reported in \cite{des-y3-3x2, des-y3-3x2-ext}.
BAO measurements have delivered spectacular results but can be cosmic variance limited where samples cover significant portions of the available volume, as is true today at the lowest redshifts ($z<0.5$). The results from DESI are expected in a year. The completion of the DESI program is likely to push the cosmic variance limited samples to $z=1.$ 

A clear lesson 
is that the experimental frontier needs to be advanced via multiple avenues, rather than focusing on only one. 
Different techniques of constraining cosmic acceleration and related phenomena are highly complementary, and the combination often enables strategies to mitigate certain types of systematic errors. An experimental program that is broad along multiple axes, including the wavelengths used for observations, cosmological length scales, and redshift ranges is our best bet for discovering new physics and avoiding being fooled by systematics that is present in any one single cosmological dataset.

Often new and exciting ways of extracting cosmological information occur with datasets of cosmic survey
experiments designed for other purposes. For example, the original Sloan Digital Sky Survey (SDSS) was
designed to measure the galaxy-galaxy power spectrum but not BAO in particular;  but the definitive first
detection of BAO is perhaps the single most important cosmological discovery that the SDSS made
\cite{astro-ph/0501171}. A more recent example is the BOSS experiment, which was never designed to measure BAO
in the cross-correlation between Lyman-$\alpha$ forest and quasars; yet this cross-correlation yields the
highest precision BAO measurement to date \cite{1311.1767}. The inverse distance ladder method of
extrapolating the value of the Hubble parameter from high to low redshift while utilizing very few assumptions
\cite{2007.08991}  was another novel combination of cosmological data that has led to one of the most
interesting cosmological tensions today \cite{2107.10291, 1710.00845}. 
We cannot forecast what novel techniques will be applied to the new, rich datasets from the proposed experiments described in this report, 
but continued theoretical, simulation and analysis work will undoubtedly make this happen.

Multi-purpose datasets should be particularly valuable given that experiments have not yielded an unambiguous
path forward for resolving the nature of cosmic acceleration. The situation is in some ways reminiscent of that in collider physics: a major motivation for the ATLAS and CMS experiments at LHC was to discover and begin to characterize the Higgs particle. Now the Higgs mass has been measured and its properties have so far been found to be compatible with the simplest consistent model, but a clear path forward such as the discovery of supersymmetry has not materialized. The community is therefore casting a wide net on the future: both with future Higgs factories as well as with other avenues, including precision measurements and various dark matter detectors. The situation with cosmology is in many respects similar. 


\subsection{Discovery Space}

There is a tremendous opportunity to make progress on these questions by pursuing large galaxy redshift
surveys. We can transfer the precision cosmological techniques now being applied at $z\lesssim 1.5$ to $2 \le
z \le 5$, reaching twice the comoving distance, less possible systematics from non-linearities,  and accessing roughly 10 times more volume. This promises potentially transformational advances on our understanding of cosmic acceleration both in the inflationary and modern eras.  



Such a program would be enabled by a new, Stage V spectroscopic facility, where we follow the Dark Energy Task Force standard for defining stages for classifying dark energy experiments \cite{Albrecht2006}.  
As was the case for previous cosmology experiments, such a facility may be designed with focused scientific goals but
the data will also enable a wide variety of guaranteed measurements and continuous results. A DESI-II program that precedes the Stage-V facility could provide a first step toward toward prototyping methods and exploring new ideas.  


\subsection{Community white papers}

Community input via the Snowmass process shows the desire for a phased program of small and large projects.
The large project is a Stage V spectroscopic facility, described also in CF3, CF5, and CF6 and the roadmap for
future facilities outlined in CF6; it would be capable of pursuing two large and complementary programs to
explore cosmic acceleration via data from the modern universe simultaneously. The Snowmass white paper
\cite{2203.07291} describes the physics that could be obtained with a lower-redshift, high density
spectroscopic survey. The higher rate of spectroscopic measurements at $z\sim 0.5$ than at $z\sim 4$ allows
construction of a high density sample providing a high-fidelity map of the cosmic web of large-scale
structure, enabling many rich statistical measurements of its properties, but demanding theoretical
development of methods to enable robust cosmological parameter estimation from non-linear scales of
overdensities. Reference \cite{2203.07506} describes the physics that could be obtained with a higher-redshift
survey tracking linear modes of overdensities. Working at $2\le z \le 6$ enables surveying of a vast volume
using only the linear modes, for which robust cosmological parameter estimations methods have been developed
in the SDSS, BOSS, eBOSS, and now DESI surveys in conjunction with the theory community. This redshift range probes the time when the universe was younger and dark matter overdensities more linear. 
These two ideas should form the backbone of future flagship experimental efforts that will continue to advance our understanding of cosmic acceleration and related phenomena. 

Community input via the Snowmass process as well as earlier DOE processes such as Cosmic Visions \cite{1604.07626,1604.07821,1802.07216} also shows a desire for smaller projects on shorter timescales, encapsulated in two broad routes towards enriching the progress of cosmology research in the next decade via studies of the moderate-redshift Universe. Blazek et al. \cite{2204.01992} describe 
great opportunities to improve the cosmological constraining power from current and near-future experiments
via modest-scale targeted efforts. Many of the potential gains would come from obtaining additional data to
complement measurements from the Rubin Observatory Legacy Survey of Space and Time (LSST). Blum et al. \cite{2203.07220}
describe the options for the use of the Rubin Observatory after the 10-year LSST is completed, for which
there are many ideas but for which there is consensus that after a few years of LSST data and science
the best way forward will become clear. Chakrabarti et al. \cite{2203.05924} describe a number of emerging new technologies that might enable precision measurements of fundamental physical observables such as redshifts and astrometric positions, for example measuring directly the $dz/dt$.

In the remainder of this report, we 
explore the opportunities to improve our understanding of cosmic acceleration and the modern universe.
In section \ref{sec:context}, we describe the current state of surveys of the modern universe. 
In section  \ref{sec:small}, we describe science opportunities that would be enabled by investments to complement near-future facilities. 
Section \ref{sec:rubin} summarizes the opportunities to make use of Rubin Observatory to study cosmology after the completion of LSST. 
Section \ref{sec:s5ss} presents the science opportunities that would be enabled by massively-multiplexed
spectroscopic capabilities, including a Stage V spectroscopic facility.
Section \ref{sec:elts} focuses specifically on those measurements which would take advantage of the extremely large telescopes that will come online over the next decade.
Finally, section \ref{sec:r_and_d} describes research, development and pathfinders for new instrumentation or methods for cosmological measurements in the future, including the  possibility of continuing observations with DESI.

\section{Context: The Experimental Landscape Today \& Tomorrow}

\label{sec:context}

The SDSS combined imaging and spectroscopic components to survey a large fraction of the extragalactic sky. In
the 2010s survey cosmology pursued by the US high energy physics community split into two natural complementary experimental tracks: imaging and spectroscopy.
Imaging experiments enabled  measurements of the matter-matter power spectra via weak lensing, one of the key
projects of both DETF Stage III  Dark Energy Survey (DES) and the Stage IV Rubin Observatory LSST.
Spectroscopic experiments enabled measurements of the galaxy-galaxy power spectra directly, especially the BAO
feature, one of the key projects of Stage III BOSS and eBOSS, and the Stage IV DESI.

In this section we will summarize the experiments to which the US high energy physics community
made large contributions. We acknowledge that this leaves out the  wider context of cosmology experiments,
such as Pan-STARRS, the Kilo-Degree Survey, and HyperSuprimeCam, but it is a decision we make for brevity.

\subsection{Stage III}

\textbf{BOSS and eBOSS}:
BOSS (the Baryon Oscillation Sky Survey) and eBOSS (the extended Baryon Oscillation Sky Survey) have been the primary Stage III spectroscopic experiments.  Both efforts have utilized an upgraded instrument, the BOSS spectrograph, on the pre-existing Sloan Digital Sky Survey telescope \cite{1208.2233}.  BOSS obtained spectra of more than 1.3 million galaxies and almost 300,000 QSOs over a sky area of more than 9000 square degrees
\cite{1208.0022,1501.00963}.  It utilized selection methods updated from those used by the original Sloan Digital Sky Survey with a goal of measuring Baryon Acoustic Oscillation scale at $z<0.7$ using Luminous Red Galaxies (LRGs) and at $z>2.1$ using the Lyman alpha forest (i.e., hydrogen gas along the line of sight to quasars).  BOSS data has also been used for a variety of other tests of cosmology, including constraints on the growth of structure from redshift-space distortions.

The eBOSS survey used the Sloan telescope, but applied target selection methods that were prototypes for the Stage IV DESI experiment \cite{2112.02026}.  In total eBOSS observed roughly 300,000 LRGs, 270,000 Emission Line Galaxies (ELGs), and more than 400,000 quasars (QSOs). 
The eBOSS survey completed its observations in 2019, and its final cosmology results, based on a wide variety of methods including BAO distance and redshift-space-distortion measurements,  were presented in papers submitted in mid-2020. The relatively short lag between these two events reflects the simpler systematics and relative simplicity and maturity of spectroscopic large-scale structure measurements; a variety of techniques have been developed and tested in the BOSS and eBOSS surveys that are now ready to be applied to future datasets.  
The final eBOSS cosmology results were presented in \cite{2007.08991}.  The results are consistent with a
simple \lcdm\ model.  If that model is extended by allowing the curvature parameter $\Omega_k$ to be varied,
the combination of CMB and eBOSS BAO measurements constrain it to be $\Omega_k = -0.001 \pm 0.0018$.
Similarly, if the equation of state parameter of dark energy $w =\frac{P}{\rho}$ is allowed to be varied, the
resulting constraint is $-1.034^{+ 0.061}_{- 0.053}$ without use of external data, consistent with the \lcdm\
value of -1.  Measurements from BOSS and eBOSS provide play a dominant role in combined Stage III constraints
(including Planck, DES, and Pantheon supernova measurements) on the cosmic matter density parameter
$\Omega_m$, the dark energy density parameter $\Omega_{\Lambda}$, the curvature parameter $\Omega_k$, the power spectrum normalization $\sigma_8$, and the Hubble parameter $H_0$, with smaller but non-negligible contributions to constraints on the equation of state parameter $w$ and the sum of neutrino masses $\Sigma m_{\nu}$ \cite{2007.08991}.

\textbf{The Dark Energy Survey:}
The Dark Energy Survey (DES) performed a 5.5 year survey of 5000 square
degree of the southern sky. The cosmology results have been published for the first three years (Y3)
of data: in these \lcdm\ remains the preferred model \cite{des-y3-3x2}, though a variety of models have been tested \cite{des-y3-3x2-ext}.
The 3x2pt analysis consists of three two-point correlation functions: 1) the angular correlation function of
lens galaxies, 2) the cross correlation of tangential shear of sources with lens galaxies, and 3) the
correlation functions of different components of ellipticities of the source galaxies. The DES has been
improving the analysis steadily; the Y3 analysis incorporates improvements to PSF modeling, treatments of
non-linear biasing and intrinsic alignments between galaxies, shear and redshift inference (including image
simulation-derived corrections for shear and for redshift biases due to blending and detection). For Y3,
redshift distributions were determined via a process that combines spectroscopic and deep multi-band
photometric redshifts from deeper DES multi-color data, cross-clustering between sources and higher quality
photo-z and spectroscopic samples, and small scale galaxy-galaxy lensing shear ratio information. The
statistical power of the Y3 data has posed unique challenges for precision cosmological inference. 

The Y3 cosmology results are summarized in \cite{des-y3-3x2}. The cosmological quantity that is best
constrained by the 3x2pt analysis is the overall amplitude of clustering in the low-z universe- $S_8$. This
allows a powerful test for consistency between growth of structure and the expansion history in the broad
class of cosmic acceleration models based on general relativity and dark energy. This test requires a CMB
anchor for matter clustering amplitude at high-z and the test becomes sharper and more general when supernova
and BAO data are used to constrain the expansion history. The DES probes matter clustering out to $z\approx 1$
so it constrains dark energy models on its own through the history of growth of structure over this $z$ range. 
The DES Y3 analysis finds that the DES results alone and in combination with eBOSS BAO \& RSD and Pantheon SN
are consistent with \lcdm\ and consistent with the CMB measurements of Planck.  The dark energy equation of state parameter $w$ is measured to be $w = -1.031\pm 0.03$ when DES is combined with BOSS and eBOSS BAO constraints \cite{1208.0022,1501.00963,2007.08991} and Pantheon Supernovae results \cite{1710.00845}. 
 \cite{des-y3-3x2-ext} explores extensions to \lcdm\ using the DES Y3 cosmology results. These  extensions
 include 1) dynamical dark energy parameterized by $w, w_a$, 2) non-zero spatial curvature, 3) varying \neff, 4) light relics varying both \neff\ and the effective mass, 5) deviations from GR parameterized by $\Sigma(k,z)$ and $\mu(k,z)$ respectively modifying lensing and Poisson equations and using RSD data, 6) variation of growth rate of structure parameterized by independent $\sigma_8$ in different redshift bins.
Less conservatively, the DES Y3 data are also consistent with other lensing and clustering results, which generally find that $S_8$ is in tension with the Planck results.

The DES also included a deep field, weekly-cadence time domain experiment aimed at cosmology with Type 1a
supernova. The 3 year data with spectroscopic redshifts of the hosts or SN was published in \cite{des-y3-sn}
and the current effort is focused on the full 6 year sample using photometric typing and redshifts. It is
notable how valuable the deep data has been for the static 3x2pt cosmology via improvements in the photometric
redshifts, in addition to providing pure expansion history measurements via SNe.

The DES is representative of the current-generation, DETF Stage-III imaging projects, including KIDS and HSC, that have established \lcdm\ as a highly effective model for describing both the expansion history and the growth of structure.

\textbf{Local Measurements of the Hubble Parameter}: The present-day value of the Hubble parameter $H_0$ may be made directly in the local universe or extrapolated from the high-redshift universe. 
One of the most reliable results on the Hubble constant came from Cepheid observations using the  Hubble Space Telescope to calibrate supernova distances. The SH0ES team \cite{shoes2022}, based on photometry of 75 Milky Way Cepheids and Gaia EDR3 parallaxes, derive $H_0 = 73.04\pm 1.04$ km s$^{-1}$ Mpc$^{-1}$.
A second local result (the Carnegie–Chicago Hubble Program, \cite{freedman2021}) uses the tip of the red giant branch to calibrate supernova distances to derive $H_0 = 69.8\pm 0.6\, ({\rm stat}) \pm 1.6\, ({\rm syst})$ km s$^{-1}$ Mpc$^{-1}$. A third, with great promise, is the ‘bright standard siren’ method applied to the gravitational wave signal from the binary neutron star merger GW170817, which  yields $H_0 = 72\pm 10$ km s$^{-1}$ Mpc$^{-1}$ when combined with two dark siren measurements \cite{1710.05835,2006.14961}. 

The cosmological measurements start with the Planck collaboration measurement of $H_0$ from the CMB \cite{1807.06209}, deriving $H_0 = 67.3 \pm 0.6$ km s$^{-1}$ Mpc$^{-1}$, which is 4$\sigma$ away from the SH0ES measurement. Also of note are the DES result using Y3 3x2 lensing and galaxy clustering anchored with BAO and big bang nucleosynthesis, to derive $H_0 = 67.6 \pm 0.9$ km s$^{-1}$ Mpc$^{-1}$ \cite{des-y3-3x2}, and the eBOSS inverse distance ladder measurement (BAO anchored supernova), to derive $H_0 = 68.2 \pm 0.8$ km s$^{-1}$ Mpc$^{-1}$ \cite{2007.08991}, and time delays in gravitational lens systems, via which \cite{2007.02941} used 40 systems to derive  $H_0 = 67.4 \pm 4.0$ km s$^{-1}$ Mpc$^{-1}$.
 
The tension between the $H_0$ values from SH0ES and high redshift calibrated $H_0$ (either from Planck under strong model priors or combination of Planck and low redshift probes in largely model independent manner) is under intense investigation for potential systematic effects and for possible new physics.  An extensive summary of the tension is provided in a Snowmass white paper \cite{2203.06142}.  Efforts are ongoing to measure $H_0$ using a wide variety of methods, including megamaser systems, gravitational wave sources used as standard sirens, and strong lens systems; these efforts should continue over the next decade. In particular, we refer to the Hubble tension as the disagreement at $4\sigma$ between the Planck collaboration value \cite{1807.06209} and the 2021 SH0ES collaboration value.

\subsection{Stage IV}

Two optical survey experiments with major involvement from the high energy physics community will deliver Stage IV-level constraints on the dark energy equation of state over the next decade; we summarize them here.

\textbf{DESI}:
The first DETF Stage IV spectroscopic Dark Energy experiment, the Dark Energy Spectroscopic Instrument (DESI) survey, is now underway.  DESI utilizes a new instrument (of the same name) installed on the 4m diameter Mayall telescope at Kitt Peak National Observatory, operated by NOIRLab.  This instrument incorporates a camera with $3^\circ$  diameter field of view, coupled to five thousand robotic fiber positioners that carry light to a set of 10 identical spectrographs covering the wavelength range from 360-980nm.  In the course of its five-year survey, which began in May 2021, DESI should measure redshifts of more than 40 million objects over a sky area of at least 14,000 square degrees, including more than 9 million bright galaxies at redshifts $z < 0.4$, more than 7 million LRGs at redshifts $0.4 < z < 1.1$, 16 million ELGs at redshifts $0.6 < z < 1.6$, and more than 2 million QSOs, most at $z > 1$.  

DESI primary probes are BAO in both galaxy clustering and the Lyman-$\alpha$ forest. DESI collaboration is working on using other techniques such as redshift-space distortions, the full shape galaxy power spectrum, the 1D and 3D Lyman-$\alpha$ forest power spectra, which have potential to significantly improve baseline results.

After one year of survey operations, DESI has already attained over 30\% of the planned sample size; it is likely that DESI will observe significantly more objects than originally planned in its fiducial survey.

\textbf{The Rubin Observatory LSST}:
Construction of the Rubin Observatory is well underway.  The ten-year LSST survey will begin once Rubin Observatory commissioning is completed, expected to be in 2024. The facility was originally proposed as the Dark Matter Telescope in 1996; it was identified as a priority for funding in both the 2008 P5 report \cite{p5_2008} and the 2010 Astronomy and Astrophysics Decadal Survey \cite{astro2010}. The community has repeatedly recognized the opportunities presented by a large collecting-area telescope with a wide field of view and large focal plane, which enables rapid surveys of the sky that are at the same time deep (due to the large mirror area and total survey time over ten years), wide (covering large fractions of the total available sky due to the high field of view), and fast (relying on short single-visit exposures combined with repeatedly returning to the same parts of the sky to enable time-domain science, including studies of supernovae and strong lens systems of import for cosmology). 

LSST will provide a foundational dataset for all areas of cosmology that require imaging (including dark
matter studies, which are described in Chapter 3 of this document). LSST will enable precision studies of Dark
Energy using five key cosmological probes: weak gravitational lensing (i.e., 3x2), strong lensing, supernovae Ia distance
measurements, large scale structure, and abundances of clusters of galaxies. LSST will enable transformative improvements in all of these areas over current-generation (Stage III) experiments such as the Dark Energy Survey, while taking advantage of the extensive experience and progress on developing methods gained from these programs. Owing to its large sky coverage and depth, LSST will also enable numerous cross-correlation studies that combine Rubin Observatory data with either spectroscopic survey samples or cosmic microwave background experiments, further strengthening constraints on cosmology. 
 
In addition to the cosmological measurements that it will provide on its own, LSST should play an essential role in future spectroscopic experiments by providing a deep and uniform catalog that will be used to select objects to observe. LSST will provide information on billions of potential targets that then will be available for spectroscopic follow-up, including Milky Way stars that may be used for dark matter studies, galaxies which trace the large-scale structure of the Universe, and transient sources of relevance for cosmology (e.g., Type Ia supernovae or optical gravitational-wave counterparts).

\subsection{Wider Context: CMB-S4 and Space-Based Surveys}

The optical surveys with substantial HEP involvement described above will be complemented by data from several other important experiments over the next decade; we summarize those projects here.  

CMB-S4 will be a ground-breaking cosmic microwave background experiment \cite{2203.08024}. It will measure fluctuations in the
cosmic microwave background (CMB) temperature and polarization both at coarse angular scales that are vital for measuring the amplitude of gravitational waves sourced by inflation and at smaller angular scales suitable for measuring the radiation content of the Universe and measuring the distribution of matter at lower redshifts (especially $z=2-4$) via gravitational lensing. CMB-S4 will enable numerous cross-correlation opportunities with lower-redshift surveys, including Sunyaev-Zeldovich maps that will enable better understanding of hot ionized gas in the universe as well as CMB lensing maps.  CMB-S4 was recommended for support in both the 2014 P5 report \cite{p5_2014} and in the 2020 Astronomy and Astrophysics Decadal Survey \cite{astro2020}.

There are also several NASA and European Space Agency (ESA) space missions which should have substantial
impact on cosmology in the coming decade. ESA's Euclid is currently scheduled to launch in 2023.  It will obtain higher-resolution optical images, lower-resolution infrared images, and low-resolution grism spectra covering a large area of sky. Euclid will be the first space-based mission focused on a cosmology survey at optical/infrared wavelengths. The data it will obtain will be highly complementary to LSST: its higher spatial resolution should aid in deblending and weak lensing calibration for LSST, whereas LSST will have better flux measurements through six different filters (versus  one optical filter for Euclid), enabling much better inference of redshifts from imaging alone. 

NASA's Nancy Grace Roman Space Telescope (NGRST) is expected to launch in 2026.  It has improved capabilities relative to Euclid due to its larger mirror and greater number of detectors, enabling better spatial resolution and improved sensitivity at infrared wavelengths. However, unlike Euclid, some NGRST time will be devoted to non-cosmology programs and competitively proposed observations.  NGRST cosmology surveys are expected to focus on smaller areas of sky than Euclid will, providing complementary information to LSST at depths better matched to what the latter telescope will deliver. 

Finally, NASA's SPHEREx mission is a smaller-scale program with a different focus. It has a considerably smaller mirror and much lower spatial resolution than either Euclid or NGRST, but it will take data using a linearly variable filter, effectively measuring low-resolution spectra in coarse pixels across the entire sky. This will enable studies of structure in the Universe at the largest scales. 
SPHEREx will measure the size of local primordial non-Gaussianity at higher precision than either Euclid or the Roman Space Telescope. 
It too will provide numerous cross-correlation opportunities with planned ground-based experiments, including LSST and DESI, which should provide much better angular (in the case of LSST) or angular and redshift (in the case of DESI) information than SPHEREx can, at the cost of either lower spectral resolution (LSST) or limited number of targets (DESI).  SPHEREx is scheduled for launch in 2024 or later.

\section{Opportunity: Cosmological Physics in the Era of DESI and LSST}

\label{sec:small}

In the coming decade, studies of cosmic acceleration will benefit from an abundance of data from two flagship experiments, DESI and LSST.  In the short term, the greatest opportunities to expand our understanding of acceleration in the modern universe will therefore come from leveraging these new capabilities in order to take maximum advantage of these new datasets.
Both enhancements to the community's ability to work collaboratively within and across experiments and the
collection of new smaller-scale datasets obtained at comparatively modest cost could significantly increase
the constraining power of the Stage IV dark energy program. The community has presented several such activities which can be undertaken over the next decade, which we describe in this section.

\subsection{Enhancing Collaborative Analyses of Stage IV Datasets}

Both the Rubin Observatory LSST and DESI will deliver exquisite data in the next decade that will allow us to
explore fundamental questions regarding the physics of the dark Universe. The datasets obtained will be the
largest and most complex of their type by a considerable margin.  They will each provide excellent constraints
on the dark energy equation of state, but the reach of the data goes far beyond that.  Additional analyses should shed light on the allowed range of models for the accelerating expansion of the universe, will provide strong constraints on the sum of all neutrino masses, and may deliver hints about the physics of the dark matter. 

At the same time these data will pose new challenges on many fronts, from modeling and simulating physics on small scales to understanding subtle systematics to assessing unexpected hints of new physics. The results from these datasets should also inform the prioritization of future experiments to improve our understanding of the dark Universe, including considerations of future usage of the Rubin Observatory. 
However, the availability of research support for extensive analyses of the complex datasets that will be obtained, incorporating the needs to address problems as they occur and to take advantage of new opportunities for scientific discovery that should arise, is less clear.  Operations funding for these experiments will ensure that the necessary data are obtained and disseminated, but does not cover the necessary cosmological analyses. Sufficient support to carry out the rich science enabled by these  datasets  will therefore be of utmost importance to the progress of our understanding of cosmic acceleration over the next decade.  

One particularly promising area where efforts are currently limited is cross-experiment analyses.  Combined analyses of multiple datasets (including all possible combinations of DESI, LSST, and CMB-S4, as well as joint analyses of LSST and space-based data) can reduce statistical uncertainties on cosmological parameters due to degeneracy-breaking from complementary probes, but also can deliver cross-correlation measurements which have reduced systematics (as any systematics not in common to both datasets will disappear in the cross-correlation) \cite{1309.5388,2203.06795}.  These opportunities are discussed more in another chapter which presents the report of the CF6 Topical Group, but should have a particularly high impact on studies of cosmic acceleration. 
Another area where collaborative efforts could be valuable is the use of LSST data for target selection for a DESI-2 survey (discussed more in \autoref{sec:desi2}).  If the proposed Northern Stripe Rubin Observatory mini-survey (as originally proposed by \cite{1904.10438}) were conducted in the first year of LSST, targets for DESI-2 could be selected at Northern declinations that can be observed more efficiently from Mayall Observatory, and the amount of area with joint LSST and DESI observations would be substantially increased.  Even in the absence of such data, early access to LSST target catalogs may facilitate DESI2 observations, which should begin around 2026.

\subsection{Enabling Science From Transient Sources}

The Rubin Observatory LSST and other projects over the next decade will find an unprecedented wealth of transient sources useful for cosmology, including both Type Ia supernovae that may be used as standard candles and gravitational wave sources that act as standard sirens.  However, LSST imaging or LIGO/Virgo/Kagara gravitational wave data alone are not sufficient to unlock the full cosmological constraining power of such objects.  As a result, comparatively small investments in follow-up imaging and especially spectroscopy can yield major improvements in our understanding of cosmic acceleration, as we describe below.  

\textbf{Follow-up observations of Type Ia supernovae:} Measurements of Type Ia supernova distances were used
in the definitive discovery of dark energy in the late 1990s for which the Nobel prize was awarded in 2011. SN are
one of the foundational probes of dark energy.
LSST will observe hundreds of thousands of SNe Ia at $z < 1$, an unprecedented sample that could be used to strongly constrain the expansion history of the universe. 
Follow-up spectroscopy from other facilities will maximize the power  of the LSST sample. This spectroscopy serves two main goals. The first is to provide spectroscopic type classifications for ``live'' SNe (i.e., while they are observable) to identify true Type Ia objects. These classifications provide training samples for the photometric sample typing algorithms used in the construction of the next generation of SN Ia cosmology samples. Even the most advanced classification techniques cannot make robust inferences without large, homogeneous and representative training sets \cite{1603.00882}. High-quality spectroscopy of individual supernovae will also enable tests of whether the distribution of properties of Type Ia supernovae in the LSST samples is evolving with redshift, providing important constraints on potential systematics in supernova cosmology such as evolutionary drifts in progenitor properties.

The second goal is to obtain spectroscopic redshifts for host galaxies of SNe that have faded away (and thus are no longer live.) The latter is not time-sensitive and can be performed opportunistically.
The LSST Deep Drilling Fields will provide the best-characterized and deepest LSST SN samples. Following the very successful experience of the OzDES survey \cite{1708.04526},  the 4MOST/TiDES program will perform long-exposure spectroscopy on all bright live SNe in the LSST Deep Drilling Fields \cite{1903.02476}, but \textit{only for the five years when 4MOST will be operational,} only half the duration of LSST.  However, for fainter and more unusual supernovae, targeted follow-up with single-target spectrographs on both moderate-aperture and large telescopes will be required; this effort will require coordination.

An efficient strategy for maximizing the size of supernova samples is to measure the redshifts for their hosts.  Table \ref{table:supernova_times} (described in more detail in \cite{2204.01992}) lists the total amount of time it would take to perform annual spectroscopy of the expected $\sim 100$ new $r < 24$ galaxy hosts of LSST supernovae per square degree spanning the five LSST deep drilling fields using different instruments.  As can be seen, the requirements are quite modest -- less than 10 nights per year with DESI would be required for this effort (though not all LSST DDFs are visible from Kitt Peak, so other facilities will be needed as well).   
The sample of hundreds of thousands of SNe that will be discovered in the main Wide/Fast/Deep LSST survey has the potential to revolutionize cosmological analyses, but its constraining power will be limited if the supernova redshifts are not accurately known. Such supernovae and their hosts could be efficiently targeted for spectroscopic observations using a subset of the fibers on a survey instrument such as DESI or a Stage V spectroscopic facility at the same time that most fibers are dedicated to other surveys (such as the ones described in \autoref{sec:s5ss}), greatly increasing the science yield from LSST supernovae.  4MOST/TiDES will obtain spectra of a substantial number of supernovae in this mode during the main 4MOST survey, but that will span only part of the duration of LSST, so additional efforts will be needed. However, such an activity would require coordination between facilities and science collaborations.  

\begin{table}[]
\centering
   \includegraphics[scale=1]{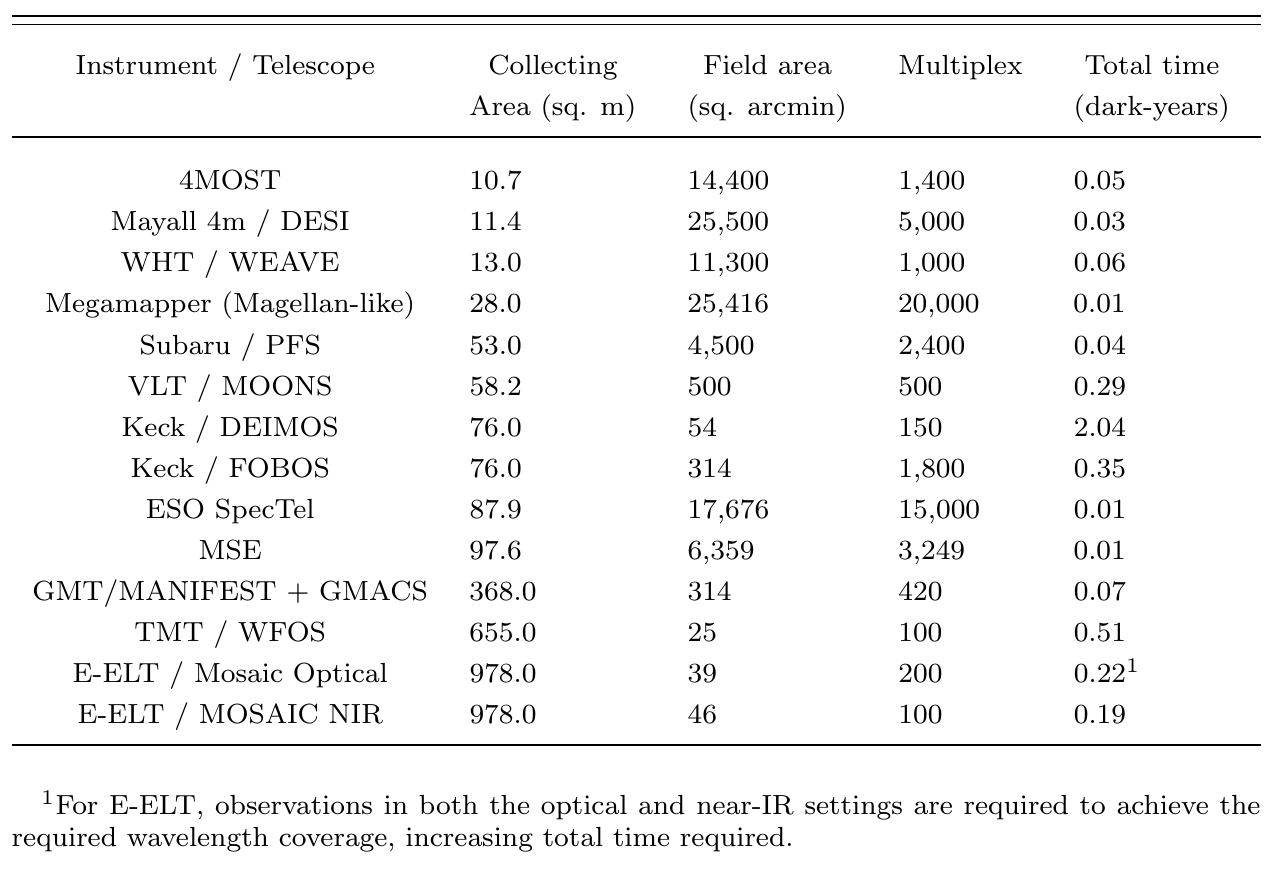}
    \caption{Time required per epoch of supernova host spectroscopy in LSST deep fields}
    \label{table:supernova_times}
\end{table}


\textbf{Using supernovae to construct the peculiar velocity power spectrum}:
Supernovae are scattered on the distance versus redshift diagram primarily due to the intrinsic variation in the luminosities of SN explosions. At low redshift or when luminosities are averaged over large samples, the scatter due to SN peculiar velocities along the line of sight can be non-negligible. These can  be useful as a cosmological probe. Peculiar velocities affect both the observed redshift and the observed flux but in a way that is not the same as transposing a supernova to a different redshift. Since peculiar velocities are correlated amongst objects that are subject to similar gravitational fields, this leads to spatially correlated residuals allowing this signal to be extracted cleanly \cite{0705.1718}. The peculiar velocity power spectrum is sensitive to the effect of gravity on the linear growth of structure. 

Measurements with high statistical confidence will be possible in the coming decade by exploiting the large numbers of low redshift ( $z <
0.1$)  SNe Ia that will be discovered in this timeframe \cite{2001.09095}; if host redshifts can be measured for large sets of LSST supernovae across the sky, measurements will also be viable at higher redshifts \cite{1008.2560}. With modest additional infrastructure compared to what is needed for LSST alone, supernova peculiar velocity measurements have the potential to provide a competitive measurements of growth at the lowest redshifts where traditional galaxy surveys are limited by the sample variance given the small total volume available.  At the lowest redshifts where supernovae are bright but few in number due to the limited volume available, it is inexpensive to obtain high-quality spectra that can improve the prediction of intrinsic supernova luminosities and hence individual distance measurements, improving constraints.   To enable the range of SN peculiar velocity science, the facilities in table~\ref{table:sn_facilities} would be needed.
The investment required would be modest, largely involving person-power for software infrastructure and operating costs, since the observations are not challenging and the instrumentation required is relatively common. 


\begin{table}[]
    \centering
    \includegraphics[scale=1]{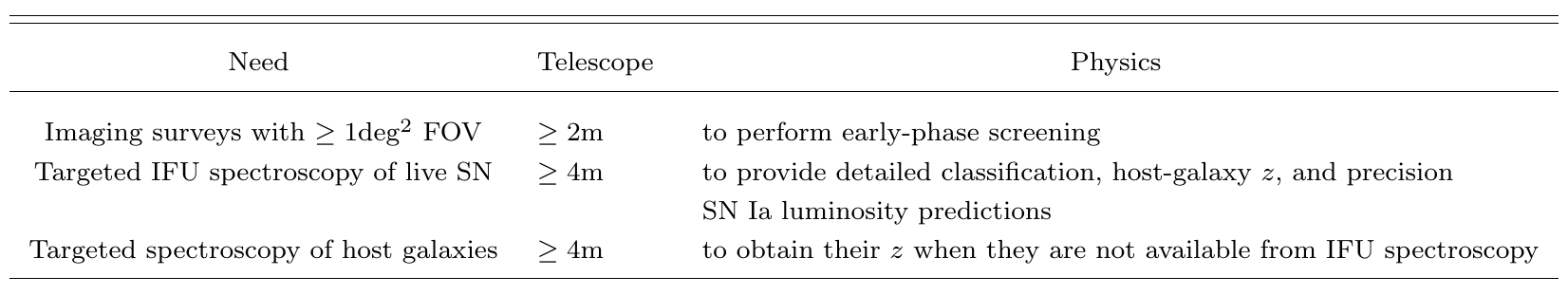}
    \caption{Facilities needed for supernova peculiar velocity power spectrum experiments}
    \label{table:sn_facilities}
\end{table}


\textbf{Exploring cosmic acceleration with standard sirens}: 
Standard-siren techniques are emerging as a very promising way to determine the Hubble parameter independent of distance ladders. Gravitational wave observatories (GWO) measure first principle distance measurements from the waveform of binary compact object merger events.  The standard-siren technique uses the GWO distance matched to a redshift measured with light. The GWO distances have well understood systematics but not very high precisions so large samples are necessary to be competitive with other methods.  The design sensitivity LIGO/Virgo/Kagara network will detect $\sim 1$ neutron star-neutron star merger per month, The next generation gravitational wave observatories will be sensitive to higher distances thus much larger volumes and thus will provide correspondingly higher rates. These samples will allow 
the distance-redshift relation to be mapped out and enable measurements of cosmic acceleration \cite{2203.11226}.  

The construction of the distance-redshift diagram 
presents two problems. The first is finding a transient associated with the merger event. As the 
the positions of gravitational wave sources are known at the $10^1-10^2$ square degree scale, there are many transients in the spatial localization volume. These are most often SN but there are varying AGN and stars. Finding the visible counterpart of a GW source therefore both requires searching a wide area of sky for transient sources, and selecting only a limited number of time-variable objects that are most likely to correspond to compact-object mergers.  The LSST Camera on the Rubin Observatory would be ideal for this work due to its large etendue and the large investment in time domain science; this could become a major mission for the observatory after the LSST is over. During LSST it will follow up some fraction of GW events; e.g., the top 10\% of events in terms of S/N and sky area localization. The standard siren technique precision gets better with number of events, however, so progress could be made by a project to follow up the top $\sim 60\%$ of events.  For example, the investment in DECam on the Blanco telescope could be leveraged by working with NOIRLab/CTIO to pay Blanco operations costs to turn it into a dedicated GW followup machine for 3-5 years. The return would be a compelling, precision measurement of the local Hubble expansion by a technique whose systematics are orthogonal to those in current use.

The second problem is measuring the spectrum of the optical counterpart or its host galaxy  to obtain its redshift.  This problem also involves making the correct identification of the counterpart in the numerous possible false positive transients, usually supernova.  There are only limited large-field, high-multiplex spectroscopic capabilities in the southern hemisphere, and they will be  saturated when event rates are as high as expected in design sensitivity LIGO/Virgo/Kagara.
When there are many potential sources distributed over a wide area of sky, the only efficient options are the widest-field-of-view multiobject spectrographs, such as DESI, 4MOST, or a Stage V spectroscopic facility.  To make a no-kilonova left behind program viable, maximizing cosmological precision, the candidate selection and SN rejection will have to be made robust, for the short run, and for the long run target-of-opportunity programs have to be implemented at wide-field multiplexed spectroscopic facilities, even those dedicated to cosmological surveys like DESI or a Stage V spectroscopic facility.  When there are few possible counterparts and search areas are small, single-object spectroscopy can be a more efficient option.

\subsection{Improving LSST Photometric Redshifts via Spectroscopy}

\label{sec:photoz}

Over the course of the ten-year Legacy Survey of Space and Time (LSST),  the Rubin Observatory will provide images of more than 20,000 square degrees of sky through six filters, providing coarse spectral information for the billions of objects detected spanning from near-ultraviolet to near-infrared wavelengths.  This coarse spectral information can be used to estimate the redshifts of individual objects or the overall redshift distribution of ensembles of galaxies; these imaging-based estimates are known as \textit{photometric redshifts} or \textit{photo-z's}.  

Photometric redshift estimates have the advantage of being available for all objects that are detectable by an
imaging survey, at the cost of lower redshift precision for individual objects. Typical photometric redshift
uncertainties $\sigma_z$ are $\sim 0.02-0.1 \times (1+z)$ in modern surveys, depending on the population studied and the dataset used \cite{2206.13633}, in comparison to uncertainties $\ll 0.001(1+z)$ from spectroscopic data. Furthermore, for a non-negligible fraction of objects photometric redshifts fail catastrophically to get the redshift correct; the fraction of objects with redshift errors $\Delta z > 0.15(1+z)$ (commonly labeled $f_{\rm outlier}$) reaches 5\% or more in deep surveys.  

Photometric redshift-dependent probes of cosmology -- which include weak-lensing shear measurements, large scale structure measurements, studies of galaxy cluster abundances, and selection of strong lens systems and supernovae for follow-up measurements \cite{1809.01669}, spanning all major probes planned for LSST --  critically depend on obtaining large samples of spectroscopic redshifts, as described extensively in prior work \cite{2206.13633,2204.01992,specneeds}.  

\begin{figure}[t]
\centering
\includegraphics[width=0.85\textwidth]{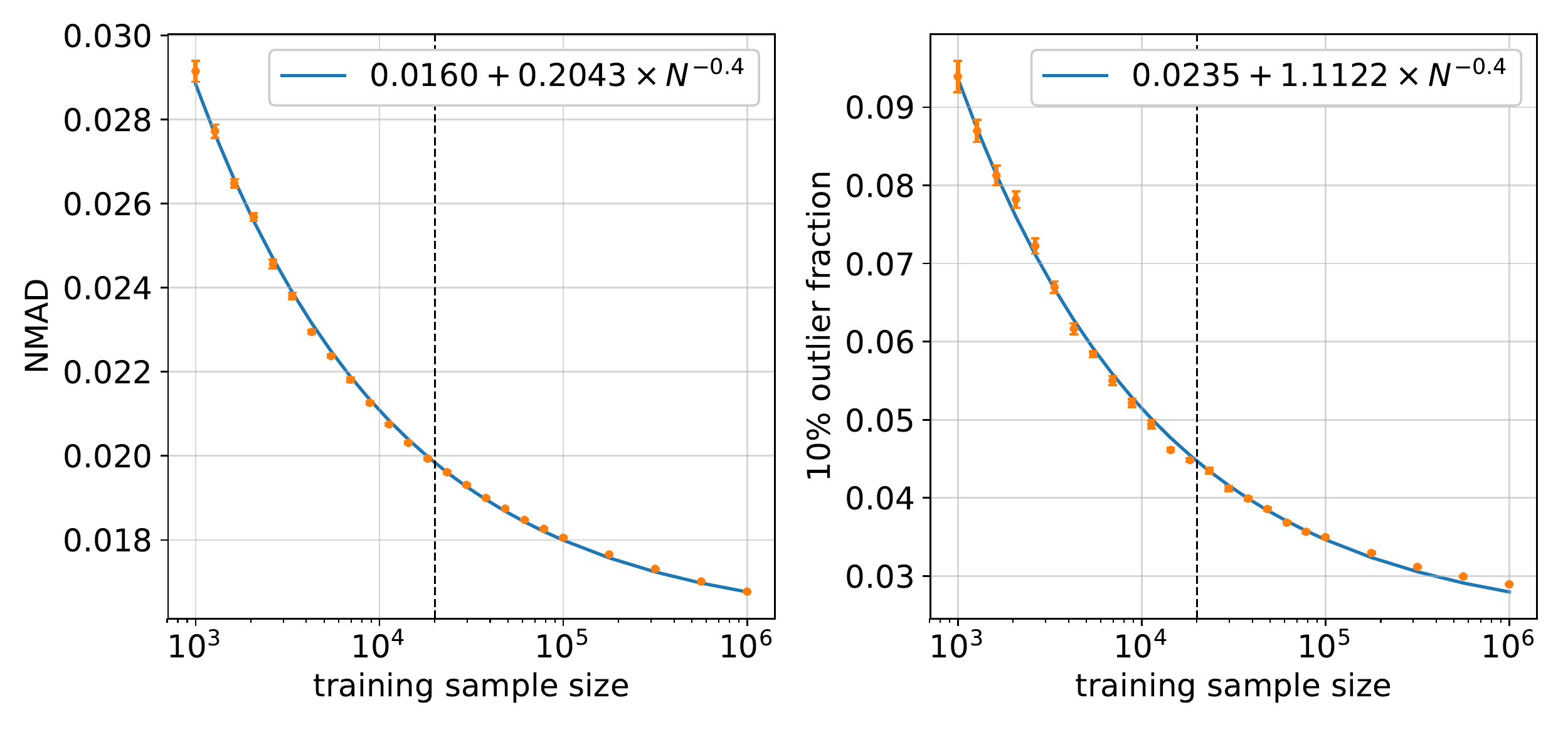}
\caption{Obtaining larger spectroscopic samples for training photometric redshift algorithms enables photometric redshift uncertainties and catastrophic outlier rates from the Rubin Observatory LSST to be reduced, significantly increasing the cosmological constraining power of this experiment. Orange points show photometric redshift errors and outlier rates versus the number of galaxies in the training set for galaxies with simulated LSST photometric errors.   Photo-$z$'s were calculated using a random forest regression algorithm. The left panel shows the photo-$z$ error, quantified by the normalized median absolute deviation (NMAD) in $(z_\text{phot}-z_\text{spec})/(1+z_\text{spec})$, as a function of training set size; similarly, the right panel shows the fraction of 10\% outliers, i.e. objects with $|z_\text{phot}-z_\text{spec}|/(1+z_\text{spec})>0.1$. A vertical dashed line shows the sample size for the baseline training survey from \cite{specneeds}.  The blue curves represent simple fits to the measurements as a function of the training set size, $N$. This analysis uses a set of simulated galaxies from Ref.~\cite{Graham2018} that spans the redshift range of $0<z<4$, using a randomly-selected testing set of $10^5$ galaxies for estimating errors and outlier rates; these catalogs are based upon simulations from Refs. \cite{grahamcat1},\cite{grahamcat2}, and \cite{grahamcat3}.}
\label{fig:training_size}
\end{figure}

\textbf{Photometric redshift training spectroscopy for LSST}: One application is to improve the \textit{performance} of photo-$z$ algorithms; i.e., 
the uncertainties in the redshifts of individual objects.  With spectroscopy of a sample of 20-30,000 objects extending as faint as the faintest objects used for LSST cosmology and distributed over multiple independent areas of sky, it should be possible to measure photometric redshifts with uncertainties approaching the expected LSST system-limited performance of $\sim 0.02(1+z)$, rather than the $\sim0.05(1+z)$ achieved in deep datasets today \cite{2204.01992,Graham2018}, as illustrated in \autoref{fig:training_size}.

This reduction in uncertainties would improve the cosmological constraining power of LSST greatly, increasing
the expected Dark Energy Task Force figure of merit from weak lensing and large scale structure measurements
alone by 40\% \cite{0806.0937}, with even larger gains to cluster cosmology likely. At the same time, the
samples of objects with spectroscopic redshifts obtained for training photo-$z$'s can be used to test and
optimize models for the intrinsic alignments between the orientations  of galaxies that are near each other in three dimensions, which represents one of the strongest astrophysical systematic effects in weak lensing analyses \cite{2204.01992}.

Such a dataset will require extensive observations to obtain; estimated survey times with different instruments and facilities are given in  \autoref{table:photoz_times}, which was originally published in \cite{2204.01992}. 
As is described below in \autoref{sec:s5ss_enabling},
a Stage V spectroscopic facility would be well-suited for this application, but no such facilities will be available until possibly late in the LSST survey. 

As a result, it would be highly valuable to undertake more limited photo-$z$ training campaigns earlier in the progress of LSST; this would both reduce the time demands on future facilities and deliver higher-quality science from Rubin Observatory sooner.  As can be seen in \autoref{table:photoz_times}, there are three modestly efficient options that would utilize 4m-class telescopes that currently exist and instruments that are operating or well underway: VISTA/4MOST, Mayall/DESI, or WHT/WEAVE.  Any of these instruments would require roughly 500 dark nights to conduct the baseline LSST photometric redshift training survey, though shallower surveys would still be useful in the shorter term and require less observing time.  Out of these, DESI is based in the U.S. and has been constructed and operated using DOE funding, making it the most promising 4m-class option to undertake this work.

The most efficient way to obtain the necessary training spectroscopy in the near future would be to utilize the PFS instrument on the 8m Subaru telescope in Hawai'i; it would require roughly 150 dark nights for the LSST baseline survey.  Subaru time is already being dedicated to observations in support of the Nancy Grace Roman Space Telescope, which should include photometric redshift training programs with PFS; the samples needed for this overlap with, but are not identical to, those needed for LSST.  A promising option would be to pursue a joint training program with NGRST using Subaru, potentially incorporating time provided to LSST as in-kind contributions as well as that provided for NGRST support.  However, significant effort would still be needed to develop target samples, conduct observing campaigns, and reduce and analyze the resulting data in order to optimize the science output from LSST.

\begin{table}[]
    \centering
    \includegraphics[scale=1]{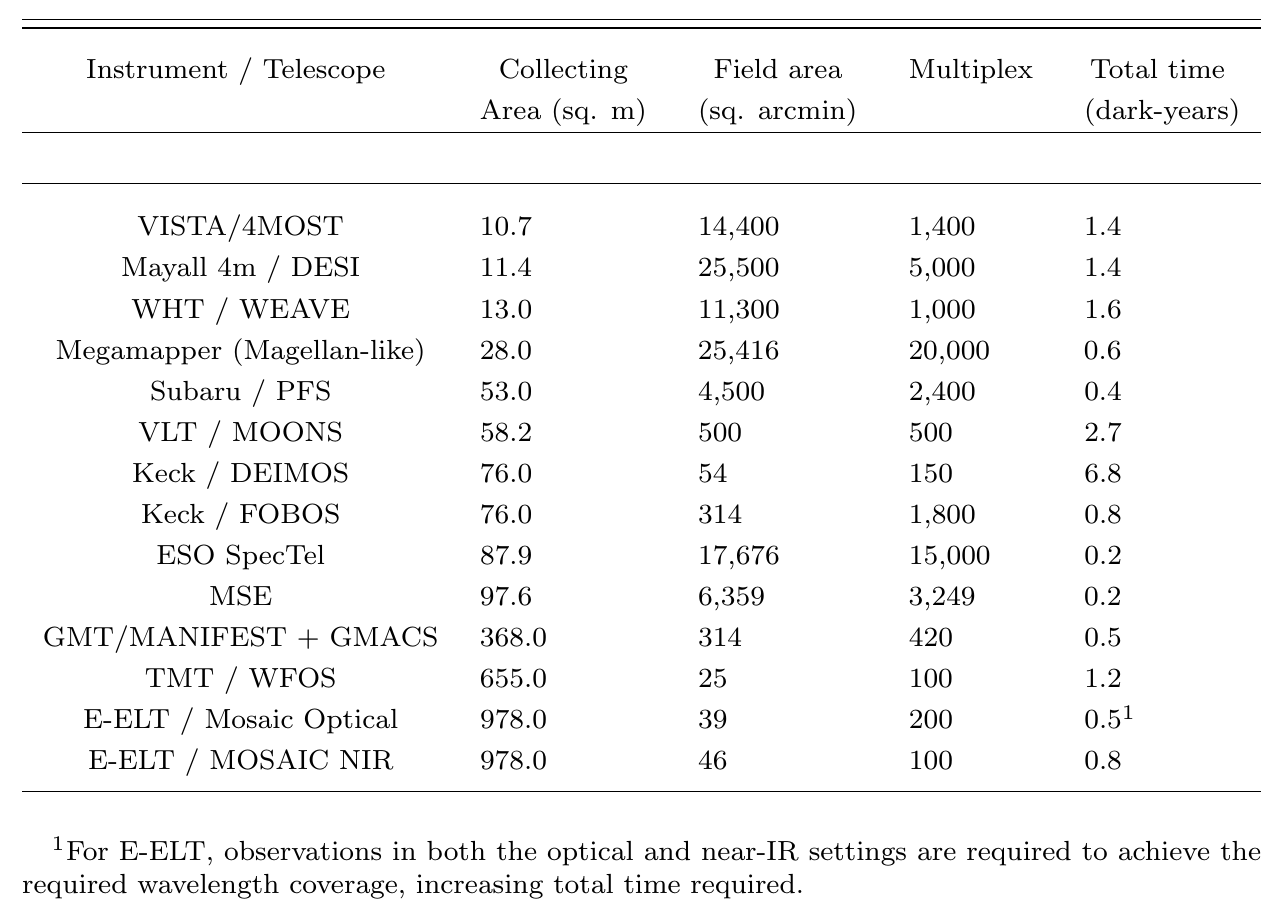}
    \caption{Time required for photometric redshift training spectroscopy}
    \label{table:photoz_times}
\end{table}


\textbf{Spectroscopy for characterizing LSST redshift distributions}:
Because of their greater uncertainties, photometric redshift-based analyses are also dependent upon having
    accurate \textit{characterization} of their error distribution (or, equivalently, of the redshift
    distributions of photometrically-selected samples).  Moments of the redshift distribution, including the
    mean and standard deviation, must be determined with exquisite accuracy (with uncertainties $\sim
    0.001(1+z)$ by the end of the survey) for LSST cosmology not to suffer systematic errors that exceed
    random uncertainties in cosmological parameters \cite{1809.01669}.  High-precision calibration of the
    redshift error distribution is also important for accurate modeling of intrinsic alignment effects
    \cite{2207.01627}. Because photo-$z$'s are intrinsically uncertain, and because photo-$z$ and intrinsic
    alignment uncertainties can mimic each other, exquisite calibration required is
    critically dependent upon obtaining spectroscopic redshift information. 

If redshifts can be robustly measured for almost all targets in the deep spectroscopic surveys that would enable optimized photo-$z$ performance, redshift distributions and errors could be characterized to the necessary precision directly \cite{2204.01992}.  However, existing deep samples have been systematically incomplete, failing to yield secure redshift measurements for $\sim 30\%$ of galaxies in a property-dependent manner.  In such a scenario, less direct methods are needed.

In that case, a promising option is to rely upon cross-correlations between photometrically-selected samples and objects of known spectroscopic redshift, as a function of the spectroscopic redshift.  The two samples of objects will cluster together when they trace the same underlying large-scale-structure -- i.e., when they overlap in redshift.  Such cross-correlation measurements, in combination with measurements of the clustering of both spectroscopic and photometric samples on their own, can provide sufficient information to reconstruct the redshift distribution of photometric samples with fidelity.  

However, such cross-correlation signals are best measured from clustering at quasi-linear or larger scales
    using spectroscopic samples that span very large areas of sky \cite{1003.0687}.  As a result, it is
    desirable to have extensive, wide-area surveys that trace the large-scale structure across the full
    redshift range spanned by LSST cosmological samples ($z \lesssim 3$).  DESI and 4MOST will provide large
    samples of galaxies at $z <\sim 1.5$ over wide areas, which should be well-suited for cross-correlation
    measurements.  At higher redshifts, these samples will only have quasars and their absorption systems as
    tracers.  These samples will map out structure only dilutely and thus in the shot noise dominated limit, and may exhibit astrophysical systematics that alter cross-correlations \cite{Gallerani2011}.  By using galaxies as tracers instead, a future survey could avoid these issues.  As a result, large, higher-redshift galaxy samples would be valuable; a Stage V spectroscopic facility could provide such samples, as described in \autoref{sec:s5ss_enabling}; a DESI-2 survey could potentially provide smaller galaxy samples at these redshifts that could enable progress in earlier years.

\section{Opportunity: The Rubin Observatory after LSST}

\label{sec:rubin}

The Vera C. Rubin Observatory will conduct LSST over 10 years beginning in 2024.  The LSST data will be the
    subject of intense analysis and provide world leading constraints on dark energy. The observatory itself,
    with a 6.5m effective diameter primary and optics providing a 10 sq-degree field of view, provides an
    etendue much larger than any other and will remain a world class facility after the completion of the survey. The Rubin Observatory after LSST presents great opportunity.

The simplest option would be to continue to operate with the same instrumentation that is used for LSST.  In that case, the primary gains for cosmology would include the continuing stream of new transients (including type Ia supernovae) that the Rubin Observatory would discover, the deeper imaging obtained from longer surveys even if the gains scale only with $\sqrt{t}$) and the potential to detect optical counterparts of gravitational wave sources through targeted follow-up imaging for which the large collecting area and field of view of Rubin provide considerable advantages for the high event rates in the next generation gravitational wave observatories.  

The next step up in complexity would be to use the LSST camera for a continuing survey, but with a different set of filters.  In particular, if a set of filters were implemented that were offset from the LSST $ugrizy$ filters by half their wavelength width, Rubin could double the amount of spectral information available for all targets by conducting a new 10-year survey. This would improve redshift accuracy for all objects by typically decreasing photometric redshift errors by 50\% from the combination of greater spectral resolution and longer total observing time.  This strategy would still maintain the same sensitivity that LSST will have for transient detection and gravitational wave source follow-up, enabling that work to continue. 

One could further increase redshift accuracy by greatly increasing the total number of filters used to $\sim 20-30$ or more (and decreasing their wavelength width accordingly), effectively obtaining a low-resolution spectrum of all objects observed.  This would be somewhat more expensive than the previous options due to the cost of Rubin filters. Narrowband filters could enable photometric redshift errors of $\sim 0.005(1+z)$ or less, potentially reaching sufficient precision to obtain radial BAO information, albeit over a small subset ($<10\%$) of the total LSST survey area. Surveys which cover the same sky area as Roman Space Telescope grism observations could be of particular interest, providing optical spectral information to complement the low-resolution infrared spectra from Roman.  Narrowband filters would yield lower sensitivity to transients than LSST, limiting the science that Rubin could undertake in this scenario.

The most expensive option and likely the most capable option would be to develop an entirely new instrument to replace the LSST camera. Among the options here are  massively-multiplexed spectroscopy, though preliminary investigations reveal significant engineering challenges.

Given that we do not yet know how Rubin Observatory and the LSST camera will perform nor what discoveries they
will make, it is premature to select one of these options now.  Instead, it would be prudent to evaluate the prospects for future science with the LSST camera as well as alternative opportunities after one to two years of LSST data have been obtained.  By that time, the prospects for a new spectroscopic facility should also be better understood, which may affect this prioritization.

\section{Opportunity: A Stage V Spectroscopic Facility}

\label{sec:s5ss}

The two Stage IV experiments will have dramatic constraining power on models of dark energy, definitively
mapping the expansion history to $z\approx 1.5$. The underlying cause of cosmic acceleration, however, may
still remain unknown, and new models may make use of dynamics of dark energy beyond the era where our
experiments have constrained it. The defining feature of \lcdm\ is that the density of dark energy is unchanged with $z$ and above $z\approx 5$ it is dynamically unimportant.
We have two possible outcomes to the Stage IV experiments: 1) the dark energy appears to be \lcdm\ to the $z$
that we probe, or 2) we find evidence for dynamical dark energy or a deviation from general relativity. In
either case it is imperative that we pursue  the program of exploring dark energy.

\begin{figure}[!h]
    \centering
    \includegraphics[width=0.8\linewidth]{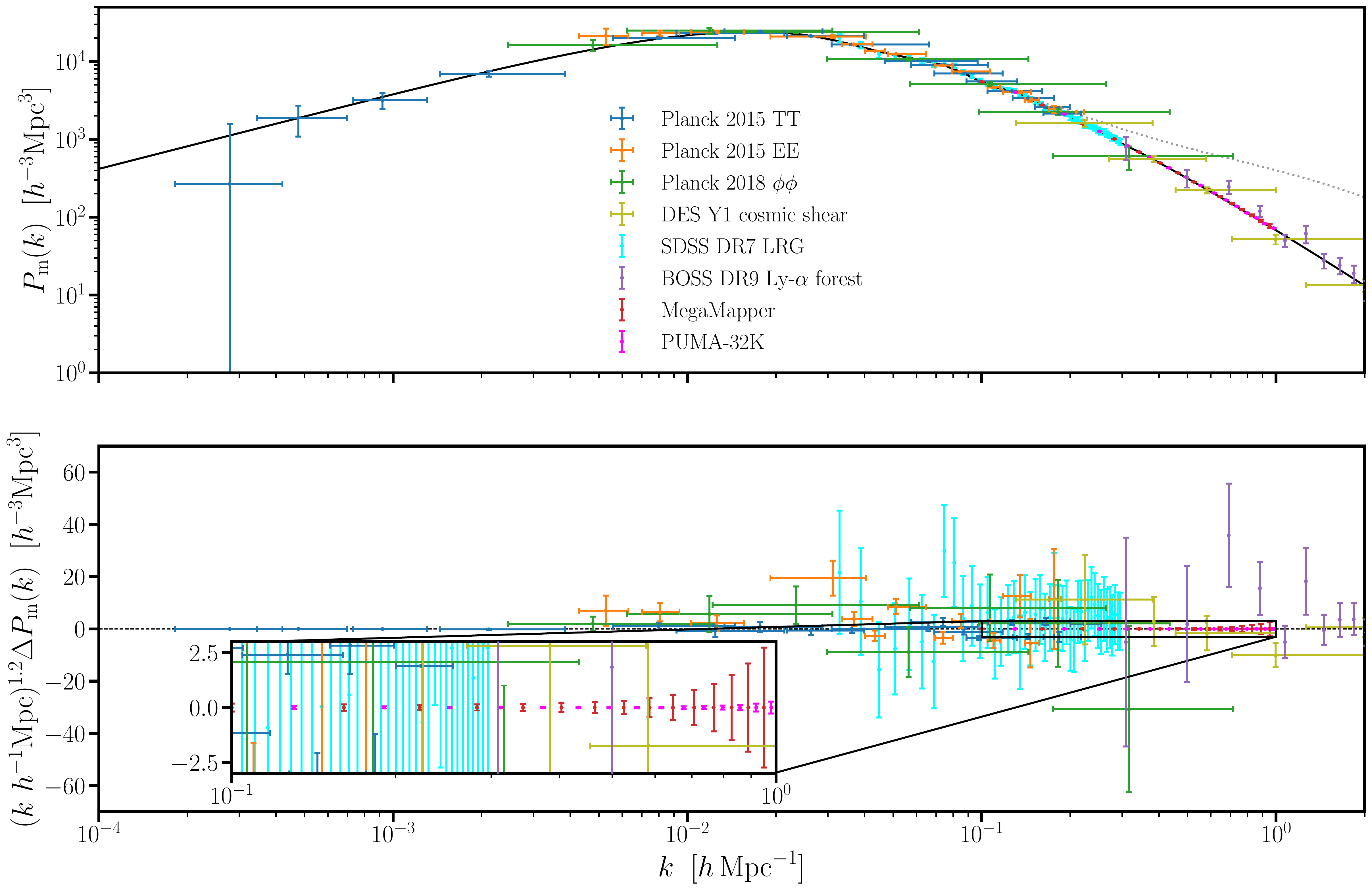}
    \caption{By observing large-scale structure in detail at higher redshifts where greater volumes of the
    Universe can be surveyed, a Stage V Spectroscopic Facility can measure the power spectrum at large scales
    with high S/N, opening new windows on inflation and other phenomena. The top panel shows previous
    measurements of the linear matter power spectrum (all normalized to $z=0$, with the nonlinear $P(k)$ as a
    dashed line), together with predicted errors for future projects. Previous constraints shown are based on Planck Cosmic Microwave Background data, weak gravitational lensing from DES, and spectroscopic samples from SDSS. Red error bars show the anticipated errors from MegaMapper -- the smallest-aperture proposal for a Stage V Spectroscopic Facility -- while purple error bars correspond to results from a PUMA-32K line intensity mapping experiment (pathfinder work that would help to explore the prospects for PUMA is described in \autoref{sec:r_and_d}).  
    The bottom panel shows the ratio of each measurement or forecast to the black \lcdm\ curve in the top panel.  This figure is reproduced from \cite{2203.07506} and adapted from refs. \cite{PlanckLegacy18, Chabanier:2019, Sailer:2021}.  } 
\label{fig:marius_plot}
\end{figure}

The next generation is a Stage V spectroscopic facility that brings 20,000--50,000 fibers to the focal plane of a 6m--10m  telescope. The massively multiplexed spectroscopy this facility allows enables
improved constraints on cosmic acceleration by making precision measurements of the matter power spectrum as
traced by multiple populations of galaxies and intergalactic gas. These precision measurements enable BAO
mapping of the distance scale out to where \lcdm\ dark energy is dynamically negligible, evaluation of the
Gaussian random field nature of galaxy spatial distributions to search for deviations from Gaussianity due to
multi-field dynamics of inflation, to put stringent constraints on all light dark radiation particles that
thermalized in the early universe, search for primordial features in the power spectrum imprinted by the
dynamics of inflation, and to  use the shape of the power spectrum to measure the sum of the neutrino masses
and place tight constraints on models where there is a third era of acceleration at $10^2 \le z \le 10^5$,
known as early dark energy. The ability of a Stage V spectroscopic facility for cosmology to make precision
measurements of the matter power spectrum at large scales is illustrated in fig~\ref{fig:marius_plot}.  The
science reach of a Stage V spectroscopic facility is immense.

A Stage V spectroscopic facility is also in an obvious follow-on to the deep imaging surveys that will be
performed over the next decade, the LSST of Rubin Observatory and the surveys of the ESA Euclid spacecraft and
the NASA Roman Space Telescope. The additional precise information on the line-of-sight direction enables
higher signal-to-noise and lower-systematics measurements of large-scale structure, but also can provide
statistical redshift information to imaging-only projects via cross-correlation measurements as a function of
spectroscopic $z$. Conversely, the imaging surveys will provide the target selection data that a Stage V spectroscopic facility requires for its experiments.

We emphasize that \textbf{a Stage V spectroscopic facility would greatly advance a wide variety of science
areas} beyond those emphasized in this report. We note in particular the cosmic studies of dark matter
described in the report of the Snowmass CF3 Topical Group (Chapter 3 of this document), but also the needs described in a wide variety of community reports both across subfields and around the world \cite{astro2020,canada_decadal, optimizing_oir,1604.07626,1610.01661,1701.01976}, 
These reports offer the potential of the US high energy physics community collaborating with other interested
groups and nations to share costs while developing the new capabilities.  Putting aside the problem of cosmic
acceleration, massively multiplexed spectroscopic telescopes \textbf{are a high priority}, as they provide
extremely flexible capabilities that will be valuable for addressing the most pressing questions of
astrophysics.  The first spectroscopic survey from the Sloan Digital Sky Survey (SDSS) provides an excellent demonstration of this.  A key goal for this survey was to measure the overall shape of the galaxy power spectrum in order to constrain the cosmic matter density $\Omega_m$ and thus resolve the key cosmology question in the mid-1990s of whether the universe was open or closed by determining whether $\Omega_m \sim 0.3$ or 1.  Within a year of the SDSS survey start, the supernova cosmology experiments had established the existence of the acceleration, providing an alternative answer to the conundrums of the 1990s. While the SDSS did measure a precise $\Omega_m$, perhaps its largest impact was the discovery of the BAO in the galaxy distribution.  We can anticipate that even if the reasons we lay out for the machine now evolve over the next decade, the need will still be high.

In the remainder of this section, we will first examine the potential of new, larger spectroscopic surveys to
constrain the nature of cosmic acceleration directly (\autoref{sec:s5ss_acceleration}); summarize other ways
in which such surveys can constrain cosmology (\autoref{sec:s5ss_cosmology}); investigate the potential to
improve constraints from Stage IV imaging surveys by using a Stage V spectroscopic facility
(\autoref{sec:s5ss_enabling}); and finally, will summarize the characteristics and status of planned and
proposed options for implementing such a survey (\autoref{sec:s5ss_facilities}).  
We focus on the general need for a Stage V spectroscopic facility rather than on a specific implementation.

\subsection{Overview of Target Samples}

\label{sec:s5ss_acceleration}

It is clear that there are two distinct regimes where it should be possible to make major advances over current experiments, based both on experience from eBOSS \& DESI and from theoretical arguments. 
One opportunity is to obtain high-density sampling of the $z<1.5$ Universe to use the extensive information on
the growth of structure over time that is potentially available at modestly nonlinear scales, with a focus on
the era when dark energy dominates the dynamics. 
The second regime is to use samples of galaxies at $2 < z < 6$) to maximize the volume over which modes in the power spectrum are still linear, enabling BAO measurements to high $z$, as well optimally studying the clustering at the very largest scales, which can be sensitive to the details of cosmic inflation in the early Universe.  

\textbf{Opportunities at $\mathbf{z < 1.5}$, high density spectroscopic surveys tracing non-linear scales:} A Snowmass white paper \cite{2203.07291} explores the case for increasing the density of samples at lower redshifts.  By obtaining redshifts for much larger samples of galaxies the sample takes advantage of the large amount of cosmological information available on smaller scales at which overdensities grow non-linearly \cite{2203.07291}. It is much easier to measure redshifts for lower-$z$ galaxies than their higher redshift counterparts, as such objects are both brighter and easier to pre-select; one can typically measure redshifts for an order of magnitude more galaxies at  $z\sim 0.5$ compared to $z\sim 4$. The proposed facilities would be capable of measuring spectra for hundreds of millions of galaxies in this redshift range. The dense sampling from such a survey would provide a high-fidelity map of the cosmic web of large-scale structure, enabling many rich statistical measurements of its properties. The main difficulty posed is one of interpretation: methods for constraining cosmological parameters from the smaller-scale, non-linear density field are still in their infancy and would need further support to be developed in full. However, it is very intuitive to expect that such a dataset would be rich in science opportunities, including ideas that have not yet been developed. For example, explanations of cosmic acceleration that rely on modified gravity may leave distinct imprints in the cosmic web that would only be detectable in such high-fidelity measurements.

The S/N of measurements of the clustering statistics improves with increasing sample density most at smaller scales.  Errors in clustering measurements improve only slowly with sample size once the product of a sample's mean number density $\bar{n}$ and power spectrum is substantially larger than \np = 1, \cite{1308.4164}.   
The DESI survey  will have sufficient density to attain this goal for modestly non-linear scales, $k \sim 0.2$\chimp, out to $z\sim1$. At smaller scales, where the clustering of dark matter should evolve non-linearly,  \np = 1 is achieved at $z < 0.4$.   This leaves substantial room for improvement by enlarging samples at modest redshifts.    

High density samples aid studies of cosmic acceleration by improving constraints on the growth of structure and redshift-space distortions. Smaller, nonlinear scales are particularly important for improving our understanding of the relationship between the observed clustering of galaxies and the underlying distribution of dark matter, important both for structure growth tests and in modified gravity tests.
Among the most important gains from a high density sample is that a broader range of galaxies will be targeted. Galaxies with different formation histories trace dark matter differently. The differences are frequently summarized by the large-scale-structure bias $b$, defined such that the observed two-point correlation function of galaxies, $\xi_g$, is equal to $b^2 \xi_m$, where $\xi_m$ is the underlying clustering of dark matter.  The mapping from the clustering of galaxies to the power spectrum of dark matter depends on $b$ (including any dependence it has on scale); so, too, does the amount of redshift-space distortion observed from a given growth rate \cite{Kaiser1987}.  As a result, being able to make the same sets of measurements using galaxy samples having very different levels of large-scale-structure bias provides vital critical consistency checks on the impact of astrophysical systematics.  Additionally, by combining clustering measurements made using differently-biased samples that are tracing the same underlying structure of dark matter, it is possible to reduce or remove the contributions of sample/cosmic variance to errors in clustering measurements, further improving constraints \cite{2009.03862}.


 Obtaining high-density spectroscopic samples at low$-z$ combines very well with large imaging-survey datasets and their associated time-domain catalogs.  For instance, such dense maps can help with determining the sources of gravitational wave detections.  Kilonovae are likely to preferentially occur in massive, early-type galaxies \cite{1608.08626}; those are precisely the sorts of galaxies that will dominate denser samples at $z<1$.  Even if the specific host galaxy of a gravitational wave source cannot be identified, the dense maps of the $z<1$ Universe would enable improved standard-siren tests on cosmology by providing measurements of the redshift distribution along the line of sight to a given event, constraining its likely redshift \cite{2006.14961}. Likewise, these samples will provide SN host galaxy spectroscopic redshifts for the large samples that will be found by the Rubin Observatory.

\textbf{Opportunities at $\mathbf{2 \lesssim z \lesssim 6}$, high-volume spectroscopic surveys tracing linear scales}:
A second Snowmass white paper \cite{2203.07506} investigated the science opportunities opened by developing new, much larger samples of $z>2$ objects with spectroscopic redshifts.  This second
approach is to focus on improving measurements of the large-scale structure traced by galaxies on large,
linear scales \cite{2203.07506}, but to extend the samples to higher z. Cosmology is best constrained at these scales when
using galaxy surveys for two reasons. First, we can model the largest scales precisely using effective field
theory (EFT) models that have well-understood convergence properties and nuisance parameters \cite{1206.2926}. Second,
the largest-scale modes that evolve only linearly retain imprints of early conditions in the universe, allowing the physics of inflation to be explored.

The Sloan Digital Sky Survey (SDSS) I and II, the BOSS survey component of SDSS-III, the eBOSS survey
component of SDSS-IV, and now the dedicated DESI experiment have measured BAO distance scale over a large
range of redshifts and theory has provided the tools to make optimal use of the linear data. The natural
extension is to apply these methods to higher z to span from $z\sim 2$ to $z\sim 6$ would quadruple the
available volume compared to current surveys, while also probing a time when the universe was younger and
hence could retain a clearer memory of their initial conditions. The data would be useful for many topics in
cosmology, including the expansion history at redshifts before dark energy domination (potentially
constraining early dark energy models), neutrino masses, the radiation content of the Universe, tests for primordial non-Gaussianity and for features in the primordial power spectrum, and beyond.

\begin{figure}[h]
    \centering
    \includegraphics[width=\linewidth]{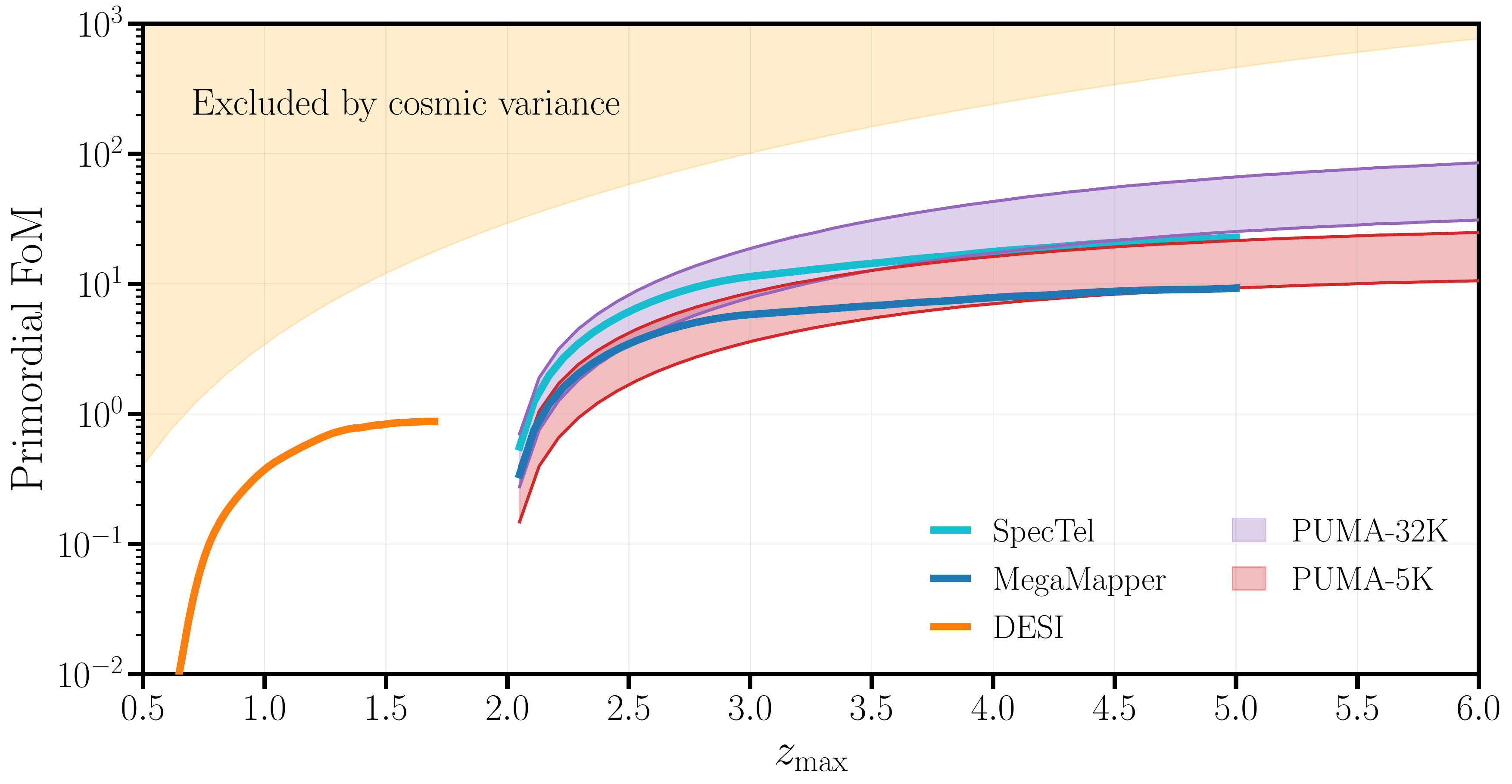}
    \caption{Due to the large volume of the Universe they would map, high-redshift samples from a Stage V Spectroscopic Facility would provide measurements of many independent power spectrum modes, opening a broad new discovery space for inflationary signatures in the matter power spectrum. The figure plots the cumulative primordial Figure of Merit (FoM), defined to be $10^{-6}\times$ the number of independent power spectrum modes surveyed, 
    as a function of maximum redshift $z_\text{max}$ for a variety of proposed samples. The orange curve corresponds to the Emission Line Galaxy sample now being observed by DESI.  Also plotted are constraints from strawman samples from the MegaMapper or SpecTel Stage V Spectroscopic concepts, as well as bands corresponding to a range of foreground estimates for the PUMA (-5K or -32K) line intensity mapping experiment,   
    The shaded orange region corresponds to FoMs larger than the cosmic variance limit for an all-sky survey, assuming galaxy bias $b(z)=1$.  Reproduced from \cite{2203.07506}.
    } 
\label{fig:figure_of_merit}
\end{figure}

The goal is to use linear modes in the power spectrum over as wide a volume as possible. Increasing $z$ provides increasing volume and the earlier times probed at higher-$z$ mean that more modes are linear. In current-generation surveys, the BAO distance scale $z>2$ has been traced only using quasars and Lyman alpha-absorbing gas along our lines of sight to them. Since quasars are rare, sample sizes have been limited by necessity, with number densities falling well short of \np$=1$ at relevant scales.  Larger samples of $z>2$ galaxies results in higher precision measurements of the BAO and thus smaller distance errors. The MegaMapper and Maunakea Spectroscopic Explorer projects have proposed to target Lyman-break galaxies \cite{Giavalisco2002} and/or Lyman-alpha emitting galaxies \cite{2012.07960} at $z>2$. Fig~\ref{fig:figure_of_merit} shows a ``primordial'' figure of merit, pFoM $\equiv 10^{-6}\,N_\text{modes}$ linear modes, for several projects.

\subsection{Stage V Physics}
\label{sec:s5ss_cosmology}

\begin{figure}[!h]
    \centering
    \includegraphics[width=0.8\linewidth]{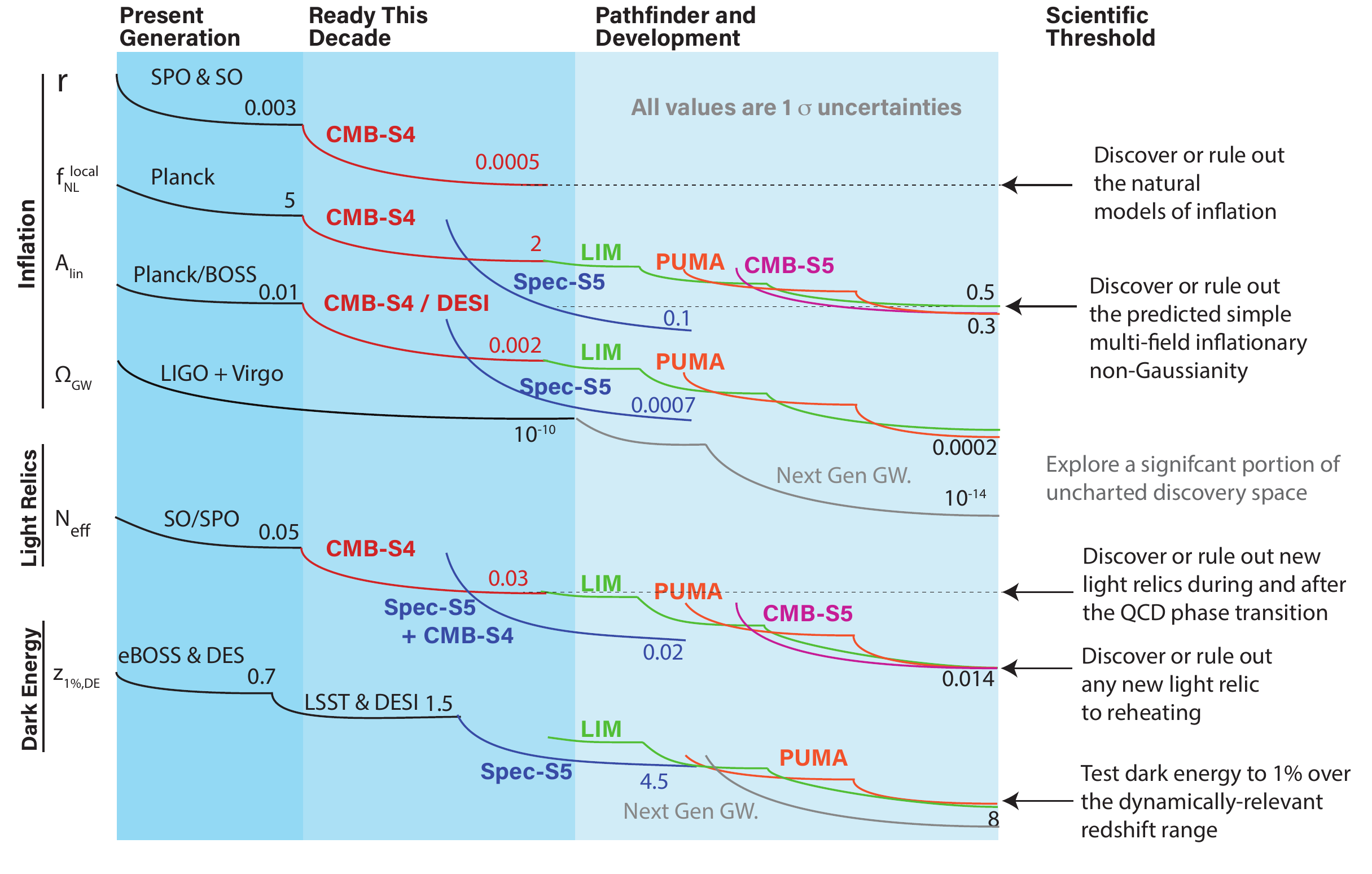}
    \caption{Illustration of  the growing scientific reach of cosmological facilities broken
into experiments that are technically ready to begin operation in this decade (2025-2035)
and more ambitious experiments requiring staged R\&D to realize facilities in the next
decade (2035- ).  This figure showcases six key cosmological observables and the associated scientific thresholds that future cosmic microwave background, spectroscopic, and line intensity mapping surveys will be able to achieve: primordial gravitational waves as parametrized by the tensor-to-scalar ratio~$r$; local primordial non-Gaussianity as measured by~\fnlloc; features in the primordial spectra parameterized by their relative linear amplitude~$A_\mathrm{lin}$; the stochastic gravitational wave background as constrained through its energy density~$\Omega_\mathrm{GW}$; dark radiation as parametrized by the effective number of relativistic degrees of freedom~\neff; and dark energy as captured by the maximum redshift~where its density is determined to better than 1\% of the total mass-energy density of the Universe, $z_{1\%,\mathrm{DE}}$.
Constraints were taken from the respective forecasting whitepapers; for the CMB-S5 we assumed   CMB-HD \cite{2203.05728}. A Spec-S5 would play a key role in searching for primordial non-Gaussianity, testing whether
    \fnlloc\ $>1$ at $>5\sigma$.  It would greatly increase sensitivity to inflationary signatures in the primordial power spectrum compared to CMB-S4, and in concert with that experiment would yield improved constraints on the effective number of light particles \neff\ (or, alternatively, provide independent constraints of comparable strength to CMB-S4). Finally, a Spec-S5 would provide precision measurements of the dark energy density parameter $\Omega_{DE}$ to much higher redshifts than ever before, $z\sim 4.5$.}
\label{fig:CF-summary}
\end{figure}

A variety of key physics goals which the community has identified for this decade, for the next decade, and
into the future are illustrated in Fig~\ref{fig:CF-summary}. The Stage V spectroscopic facility would
significantly advance our understanding of all but the CMB-specific B-mode detection of the tensor-to-scalar
ratio $r$ and the gravitational-wave-specific detection of $\Omega_{GW}$. In fact, of the 5 areas of beyond
the standard model physics for which there is strong evidence -- 
dark energy, dark matter, early causal density fluctuations (\ie\ what inflation solves), neutrino masses, and the baryon asymmetry -- only the last would not be substantially addressed by a Stage V spectroscopic facility.
Even beyond this range of measurements, it is important to note that the \textit{same datasets} will enable a
variety of other tests of cosmology; based on past experience, it is also likely that new, valuable methods
for cosmological tests enabled by the rich Spec-S5 dataset will be developed in the coming years by theorists, observers and data scientists. 
We note that although \autoref{fig:CF-summary} summarizes the CF4 Late Universe and CF5 Early Universe measurements, a Stage V spectroscopic facility would also play a major role in our advancing understanding of dark matter, as laid out in the chapter of this report on Cosmic Probes of Dark Matter (CF3).

\subsubsection{Key Quantitative Targets}

A Stage V spectroscopic facility would be able to achieve key, transformative quantitative thresholds on multiple important physics goals. 
In this subsection, we describe how Spec-S5 would address two 
fundamentally important problems with theoretically-motivated, quantitative targets, by providing precision
tests of dark energy models at the highest redshifts feasible from optical spectroscopy, and by providing tests for non-Gaussianity in cosmic structure with sufficient statistical precision to rule out many inflation models.

\textbf{Measure dark energy out to $\mathbf{z=5}$ where $\mathbf{\Lambda}$CDM becomes dynamically negligible}

\begin{figure}[!h]
    \centering
    \includegraphics[width=1\linewidth]{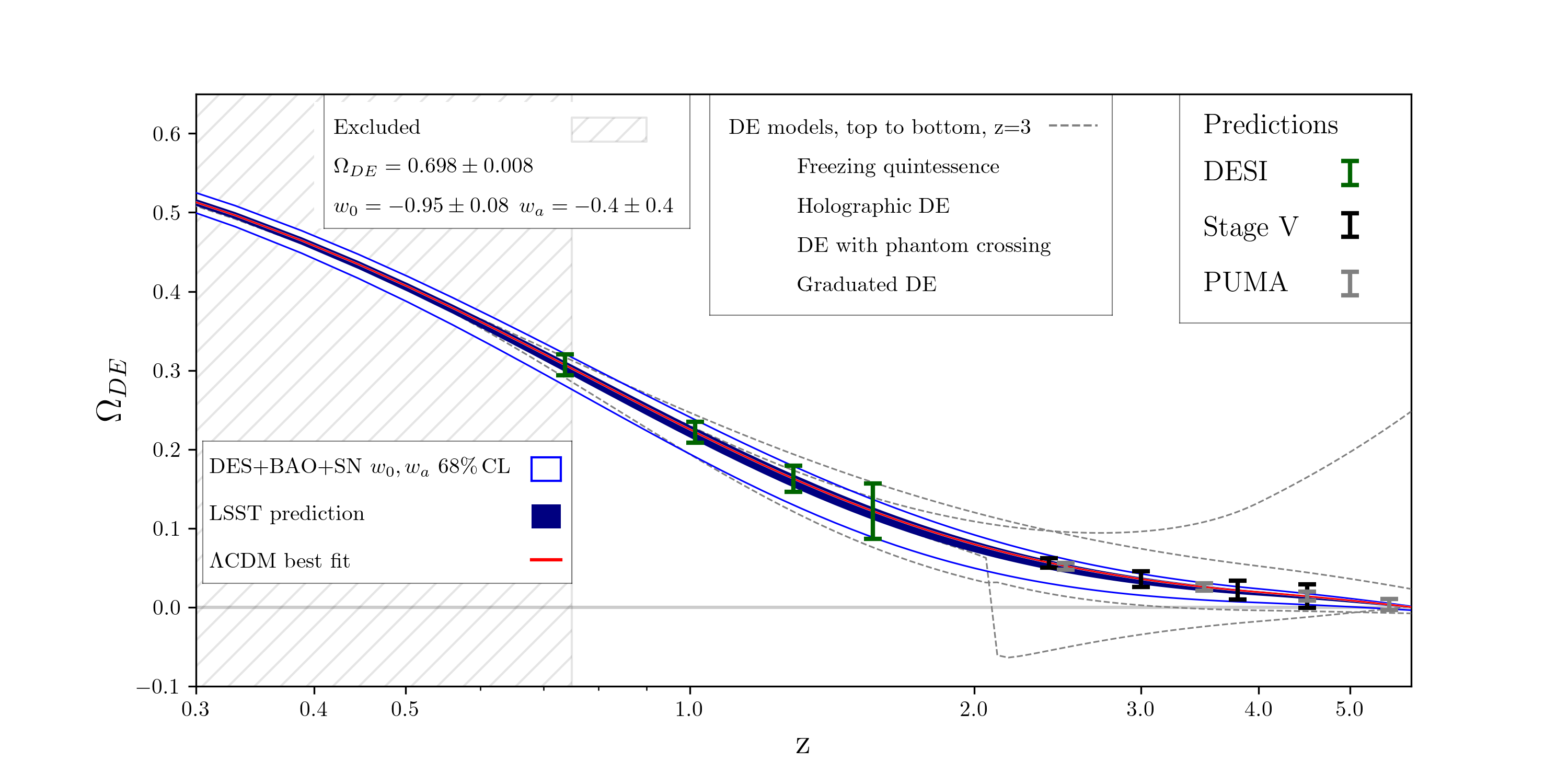}
    \caption{A Stage V Spectroscopic Facility would provide accurate measurements of the energy density of
    dark energy at $1.5 < z < 5$, a critical regime for testing alternative models which have been invoked to
    reduce tensions in other datasets as \lcdm\ has negligible density above $z=5$. Solid curves show the
    allowed regions for both a standard \lcdm\  model (red) and a  model with evolving equation-of-state
    parameters $w_0$ \& $w_a$ (blue), based on current Stage III experiment constraints from
    \cite{des-y3-3x2-ext} (best-fit parameters are $\Omega_{DE} = 0.698\pm 0.008$, $w_0 = -0.95 \pm 0.08$,
    $w_a = -0.4 \pm 0.4$). The data used in these fits are almost entirely at $0.1 < z < 0.75$, but the Chevallier-Polarski-Linder (CPL)
    parameterization of equation-of-state evolution is sufficiently constrained that it allows little
    variation in behavior at higher $z$. The \lcdm\ predictions for the DESI ELG sample, high-redshift samples
    from a Stage V Spectroscopic Facility, and line intensity mapping from the 21cm experiment PUMA are shown
    as data points with error bars (based on predictions from \cite{2203.07506}).     Predictions for a
    variety of alternative dark energy models from the literature, most of which have been proposed to
    address the Hubble tension, are also shown. 
    For a more detailed study of constraints on DE models, see \cite{2007.02865}. The dark energy models plotted from top to bottom (based on their ordering at $z=3$), are freezing quintessence with $w=-1+-8.5(\frac{z}{1+z})^7$ \cite{freezingq2016}; holographic DE with C=0.72 \cite{holographic2017,holographic2020}; DE with phantom crossing with $z_m=0.18, \alpha=1, \beta=6$ \cite{phantom2021}; and graduated DE with $\gamma=0.014, \lambda=20$ \cite{graduated-dark-energy}.  Spec-S5 would deliver sufficiently precise measurements to test CPL and its alternatives at high redshifts where they differ most.
    } 
\label{fig:ode}
\end{figure}
    
The study of dark energy has proceeded by measurements of both the expansion history and the growth of structure via the evolution of the matter power spectrum.
A Stage V spectroscopic facility is capable of measurements of the expansion history via the BAO distance scale at $2<z<5$ with $\sim 0.5\%$ errors in four separate redshift bins, a factor of 5 reduction in errors compared with the $\sim 2.5\%$ errors expected from DESI quasars and Lyman-alpha forest \cite{1611.00036}.  
A Stage V spectroscopic facility measurement of redshift-space distortions at the same $z$ range, in combination with the BAO measurements, allows $\Omega_{DE}(z)$ to be measured with
$<2\%$ errors at $2<z<5$ constraining the space of possible models for cosmic acceleration significantly, as shown in Fig~\ref{fig:ode}.


\begin{figure}[!h]
\centering
\includegraphics[scale=0.3]{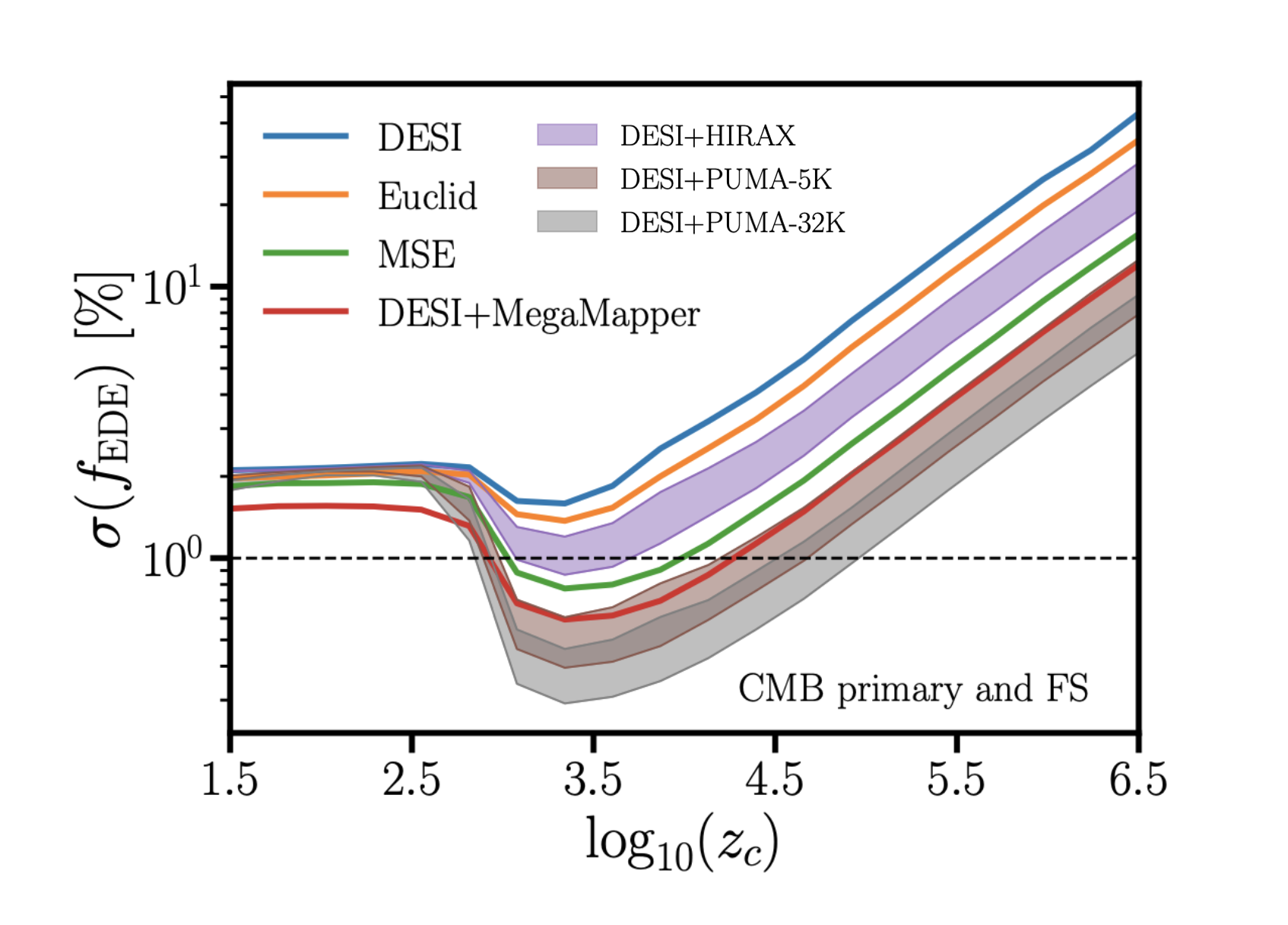}
\caption{A Stage V Spectroscopic Facility can provide sensitive tests for dark-energy-like components of the
    Universe that caused acceleration at early times ($10^2 < z < 10^5$). Plotted are predicted uncertainties
    on the maximum amplitude of early dark energy ($f_\text{EDE}$) as a function of the redshift at which EDE
    peaks $z_c$. 
    A Planck+Simons Observatory prior on \lcdm\ is included for all
    experiments. The Maunakea Spectroscopic Explorer (MSE) and MegaMapper are both proposed concepts for
    Spec-S5. The ``CMB primary and FS`` in the figure refers to constraints from combined CMB and
    spectroscopic full spectrum shape measurements. Constraints using HIRAX and PUMA are shown in bands
    because of the uncertainty in foreground removal.
Reproduced from \cite{2203.07506}. 
}
\label{fig:EDE_constraints}
\end{figure}

In \lcdm, dark energy behaves as a cosmological constant with fixed energy density at all redshifts.  In such
scenarios its contribution to the mass-energy density of the universe remains subdominant until late times
(after the density of matter and radiation has decreased due to the Universe's expansion); dark energy
represents almost 70\% of the mass-energy density today, but would be $\sim 7.5\%$ at $z=2$ and  $\sim 1\%$ at
$z=5$.  Simple models parameterized by $w=w_0+w_a\frac{z}{1+z}$ behave similarly, where $w$ is the equation of
state of the dark energy in cosmological dynamics. The current constraints on dark energy from DES Y3 3x2 galaxy and shear correlation functions, the eBOSS DR16 BAO and RSD measurements, and the Pantheon supernova samples \cite{des-y3-3x2-ext}
($\Omega_{DE} = 0.698\pm 0.008,\, w_o = -0.95 \pm 0.08,\, w_a = -0.4 \pm 0.4$) follow a \lcdm\ model closely, but most of the constraining power is derived from data at $z\lesssim 3/4$.

This is important because the space of dark energy models is not well captured by the $w_0,w_a$
parameterization and extrapolating to higher $z$ using it is unsupported. There are classes of models known as
``tracker'' models (also known as freezing quintessence) that asymptote to $w=-1$ at low-$z$ but have
significantly different values of $w$ at $z>2$  \cite{2007.02865}. Many of the dark energy models proposed to
solve the Hubble Tension (see \eg\ \cite{2203.06142}) involve phase transitions, step functions, or
oscillatory behavior of the dark energy density at $z>2$ and the most successful of these, early dark energy,
invokes a second era of dynamically important dark energy at $10^2<z<10^5$. 

As a result, the measurement of $\Omega_{DE}$ at $z=2-5$ will rule out a variety of models (see Fig~\ref{fig:EDE_constraints}).  A Stage V spectroscopic facility can determine $\Omega_{DE}$ with uncertainties of $< 0.02$ (i.e. to constrain them to better than 2\% of overall energy density of the universe at the time) across this redshift range via the combination of BAO and redshift-space distortion measurements. 
Both the dense low-$z$ and the high-volume high-$z$ surveys combine to make the total power spectrum shape
measurements needed to constrain the early dark energy models to better than $1\%$ at redshifts spanning $500 \lesssim z \lesssim  10^4$ \cite{2203.07506}. 

\textbf{Test signatures of non-Gaussianity from inflation at a sufficient precision to explore all non-fine-tuned multi-field models}

The nature of the field which led to inflation in the early universe remains a key open question in high energy physics, with only limited means to explore it \cite{2203.08128}.  Current observations of primordial fluctuations both in the CMB and in the positions of galaxies are consistent with Gaussian statistics. At the same time, deviations from Gaussianity are necessarily present even in the simplest models of inflation. More generally, primordial non-Gaussianity (PNG) is a robust probe of interaction dynamics during inflation beyond the free propagation of curvature fluctuations. Detecting and characterizing PNG would be a fantastic triumph of theoretical and observational cosmology, probing the dynamics of the early universe and providing clues about physics at very high energy densities, much higher than those achievable in particle colliders.

\begin{figure}[!h]
    \centering
    \includegraphics{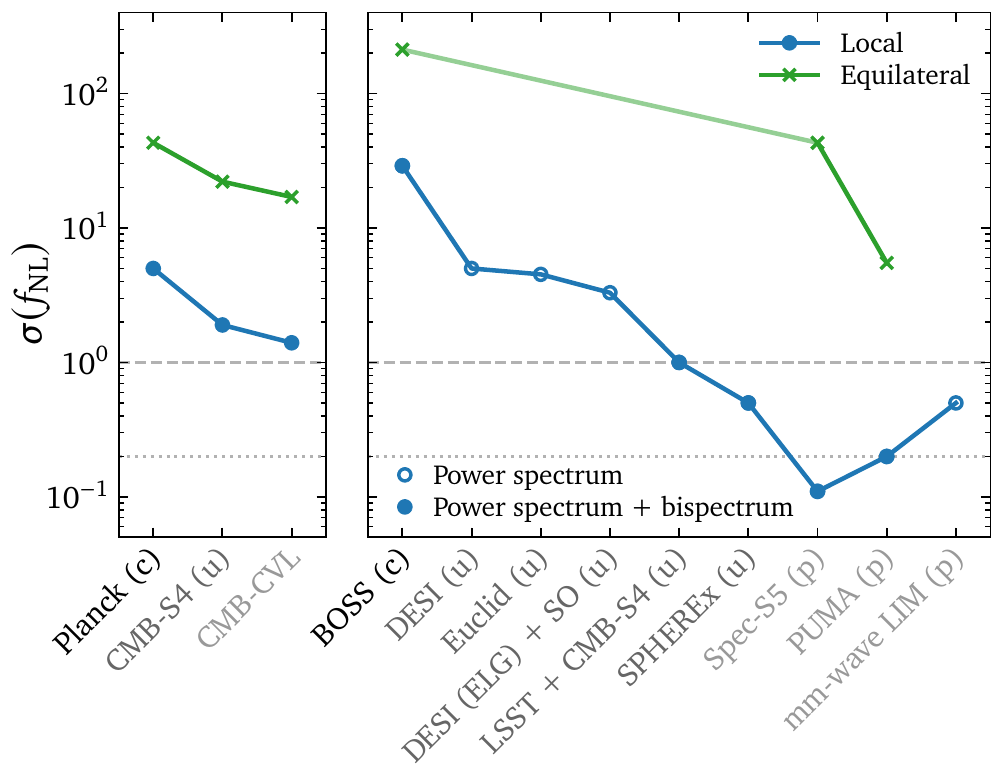}
    \caption{A Stage V Spectroscopic Facility would have greater capability to constrain signatures of
    primordial non-Gaussianity from inflation than any other project that is technically ready to proceed.
    Plotted are projected constraints on two types of primordial non-Gaussianity for CMB measurements (left)
    and late universe measurements (right) from a small subset of completed~(`c'), upcoming~(`u'), or
    proposed~(`p') experiments~(see~\cite{ Sailer:2021, Mueller:2021tqa, Castorina:2019wmr, SPHEREx, Munchmeyer:2018eey,
    Planck:2019kim, Ferraro:2019uce, PUMA:2019jwd, Abazajian:2019eic, 
    2201.11518, MoradinezhadDizgah:2018lac, Chen:2021vba, 2203.07258} for the underlying data analyses and
    forecasts). The open dots marked power spectrum only use scale-dependent bias to measure the local
    non-Gaussianity parameter \fnlloc\  while the filled dots include bispectrum information. While methods
    for projecting limits on  \fnlloc\ are well-developed, projections for equilateral \fnl\ are expected to
    improve over time with better theoretical modeling. Measuring these quantities at the \fnl\ $\approx 1$
    level would have great value: for the local \fnl\ this constrains many classes of multifield inflation
    models, while an equilateral \fnl\ signal at this level distinguishes slow roll from other mechanisms for
    single-field inflation.  Figure adapted from \cite{2203.08128}.
    } 
\label{fig:fnl}
\end{figure}

PNG can be broken into classes, and observationally the most interesting is the  local PNG, which corresponds to a bispectrum in the squeezed triangle configuration with one short side and two long ones. Intuitively it can be pictured as a correlation between the local amplitude of a small scale power spectrum and the value of the large scale density fluctuation in the same region. 

A large local PNG indicates the presence of more than one light field during inflation. Physically, the fluctuations of massless scalars freeze after horizon exit and open up a multi-field space for superhorizon evolution. Patches of the universe of Hubble size evolve independently of each other, leading to non-Gaussianities that are local in these Hubble patches. This gives rise to PNG that is local in real space. 

Being able to detect PNG at the level of \fnlloc\ $\sim 1$ is a natural target, as in non-tuned multi-field inflation scenarios its value 
its value is generically larger than unity. Therefore, \textbf{by excluding}  $\mathbf{f_{nl}^{loc} \sim 1}$
\textbf{we can exclude any non fine-tuned multi-field inflation model.}

\begin{table}[]
    \centering
    \includegraphics[scale=1]{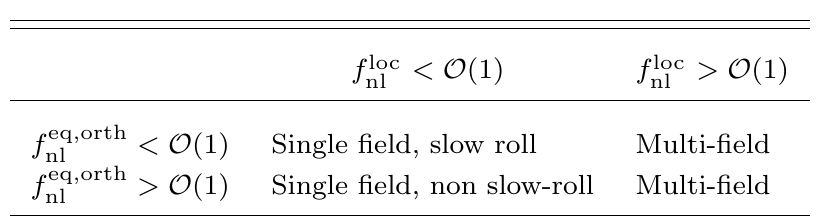}
    \caption{Constraints on inflationary mechanisms from measurements of different types of primordial non-Gaussianity.}
    \label{table:non-gauss}
\end{table}


\fnlloc\ can be measured from galaxy surveys by two means. The first is as a scale dependence in the apparent bias between the clustering of galaxies and of matter at the very largest scales.  The second is through direct measurements of its impact on the galaxy bispectrum. 
Both of these avenues have been implemented on SDSS/BOSS data, cf. e.g. \cite{0805.3580, Mueller:2021tqa, Castorina:2019wmr} 
and \cite{2201.11518, 2204.01781}, respectively.
Detecting these signals is best done at high redshifts where the number of modes with large spatial scales will be greatest due to the larger volume available per unit $z$.  

A Stage V spectroscopic facility should be able to measure the signatures of local PNG with sufficiently small
statistical errors to be able to detect a signal at the level of \fnlloc\ $=1$ at $>5\sigma$ significance (cf. \autoref{fig:fnl}). 
Due to the three-dimensional information provided by spectroscopy and the low cosmic variance due to the large volume surveyed at high redshifts, proposed high-$z$ spectroscopic surveys should yield stronger constraints on \fnlloc\ than can be achieved from CMB measurements.

\subsubsection{Opening Up New Discovery Space}

The second class of problems are those where important targets associated with important theoretical milestones are either elusive or probably beyond the reach of the next-generation experiments.  A Stage V spectroscopic facility offers great opportunities to opening up new discovery space in our understanding of several important problems of current interest in this class.

\textbf{Limits on dark radiation: constraints on all sub-eV particles thermalized in the Universe's history}

\begin{figure}[!h]
    \centering
    \includegraphics{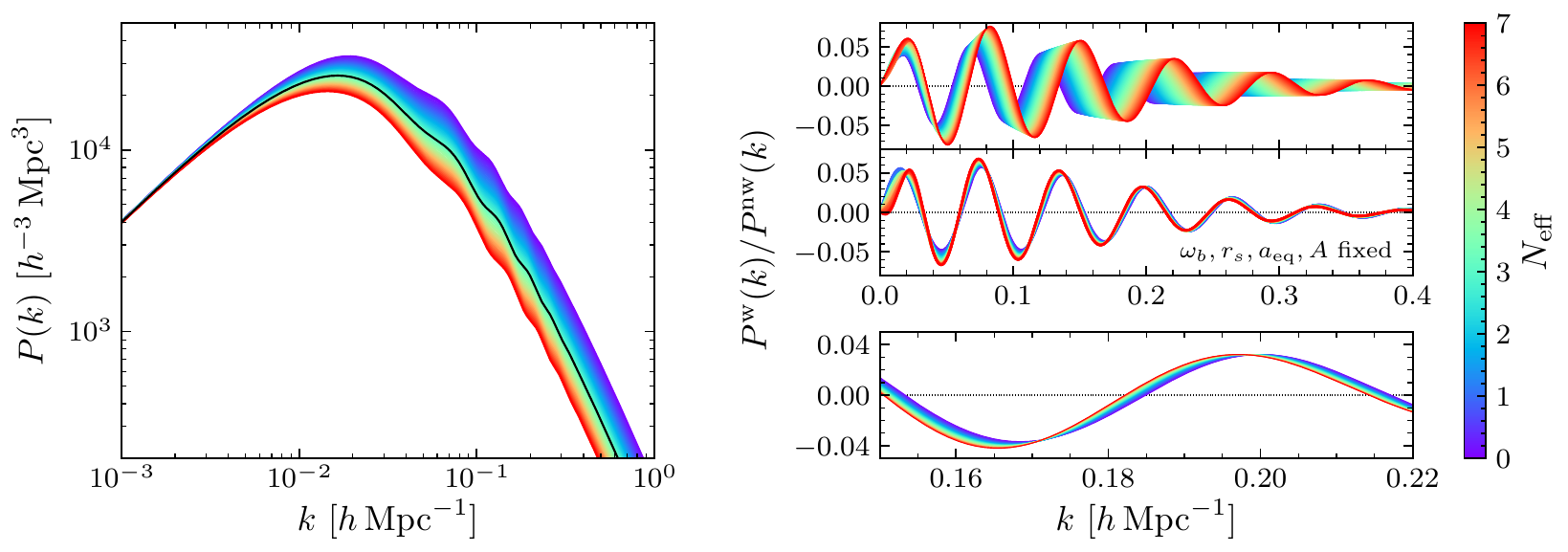}
    \caption{
Detailed measurements of the matter power spectrum, which can be traced by galaxies in Spec-S5 survey samples, can provide a sensitive probe of the effective number of relativistic species, \neff. Panels show the variation of the matter power spectrum~$P(k)$~(\textit{left panel}) and the BAO~spectrum~$P^{\rm w}(k)/P^{\rm nw}(k)$~(\textit{right panels}) as \neff\ changes (indicated by varying colors). The BAO~spectrum, corresponding to the ratio of the oscillatory part~$P^{\rm w}(k)$ of the matter power spectrum to its smooth broadband part~$P^{\rm nw}(k) = P(k) - P^{\rm w}(k)$, is shown at top right, color-coded according to the value of \neff. The physical baryon density,~$\omega_b$, the physical sound horizon at the drag epoch,~$r_s$, the scale factor at matter-radiation equality, $a_{\rm eq}$, and the BAO~amplitude~$A$ at the fourth peak are all held fixed in the second BAO~panel. This panel and the bottom zoom-in show the remaining phase shift induced by varying numbers of free-streaming relativistic species We refer to~\cite{Wallisch:2018rzj} for additional details.  Figure taken from~\cite{Abazajian:2022ofy} which was slightly adapted from~\cite{Wallisch:2018rzj}; caption adapted from~\cite{Abazajian:2022ofy}.
} 
\label{fig:neff_effect}
\end{figure}

All sub-eV particles in thermal equilibrium in the early Universe contribute to the expansion history in a
well defined and measurable way and affects the shape of the matter power spectrum and the BAO (see \autoref{fig:neff_effect}).

\begin{figure}[!h]
    \centering
    \includegraphics{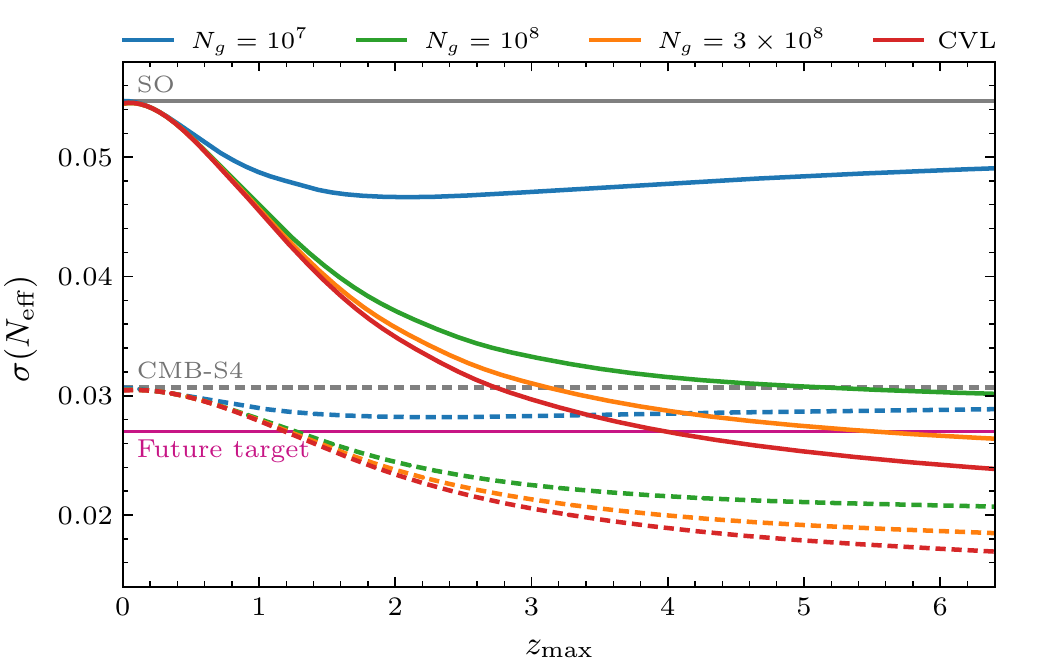}
    \caption{
Power spectrum measurements from a Stage V spectroscopic facility would greatly increase the sensitivity to
    new relativistic species compared to CMB data alone.  This figure shows uncertainties in the effective
    number of relativistic species, \neff, determined from the full galaxy power spectrum up to
    $k_{\rm max} = 0.2h$Mpc$^{-1}$ as a function of the maximum redshift surveyed, $z_{\rm max}$.
    Different curves correspond to different total numbers of objects with spectroscopic redshift
    measurements, $N_g$, at fixed survey area $\Omega = 20,000$ deg$^2$.  The comoving number density is
    assumed to be constant and fixed by the total volume of the survey within $z_{\rm max}$. For “CVL” (red),
    all modes in the survey are assumed to be measured up to the limit set by cosmic variance. Solid and
    dashed lines correspond to combining the LSS data with Simons Observatory (representative of CMB Stage
    III) or CMB-S4 data, respectively. The gray lines indicate the level of sensitivity of the respective CMB
    experiments alone. A new non-standard model Goldstone boson, Weyl fermion, or vector boson would alter \neff\ by a minimum of 0.054, 0.047, or 0.027, respectively (depending upon the temperature at which it freezes out); the latter limit is shown as a horizontal line (labelled ``Future target'').
Figure and caption adapted from \cite{baumann2018}.
} 
\label{fig:neff}
\end{figure}

The early Universe was filled with a hot dense plasma and underwent a very rapid, radiation-dominated expansion.  The extremely high temperatures in the early universe suggest that even very feebly interacting particles coupled to standard model degrees of freedom would have been at sufficiently early times in equilibrium and thus would have been abundantly produced in the early Universe. 
Particles once in thermal equilibrium and stable on cosmological time scales persist as thermal relics. Massive thermal relics became non-relativistic at early times and provide candidates for cold dark matter. Light thermal relics remained relativistic throughout the early universe, contributing to the radiation density during this epoch. 
The relic density of any new light species that was ever in thermal equilibrium with the SM plasma is given by $\Delta N_\mathrm{eff} \geq 0.027$, where the effective number of relativistic degrees of freedom $N_\mathrm{eff} = N_\mathrm{eff}^\mathrm{SM} + \Delta N_\mathrm{eff}$ parametrizes the non-photon radiation density in the Universe in terms of the energy density of one neutrino, with $N_\mathrm{eff}^\mathrm{SM} = 3.044$ from SM neutrinos.
Measurements of the light relic density therefore serve as an extremely broad probe of new physics. \autoref{fig:neff} shows how the redshift coverage and sample size of a Spec-S5 survey affects the constraint.
See \cite{2203.07943} for a more complete description.

This important measurement is given increased importance by the vibrancy of dark complexity in current theoretical discourse.  The idea is that the standard model of particle physics is very far from minimal, and we may not have predicted its complexity from first principles. As we know little of the physics of the dark matter making up $25\%$ of the Universe's energy budget today, but very much about the complexity of the $5\%$ normal matter contribution, we can ask if there is also complexity in the dark matter sector. This bottom-up plausibility argument sets the stage for considering dark complexity, the possibility of hidden sectors with particles and forces that approach the SM in complexity, with possibly equally interesting dynamics in the early universe and today. Placing stringent limits on  \neff\ places constraints on entire sectors of dark radiation models.
See \cite{2203.07943} for a more complete description.

\textbf{Search for inflationary primordial features that signal the breaking of scale invariance}

\begin{figure}[!h]
    \centering
    \includegraphics{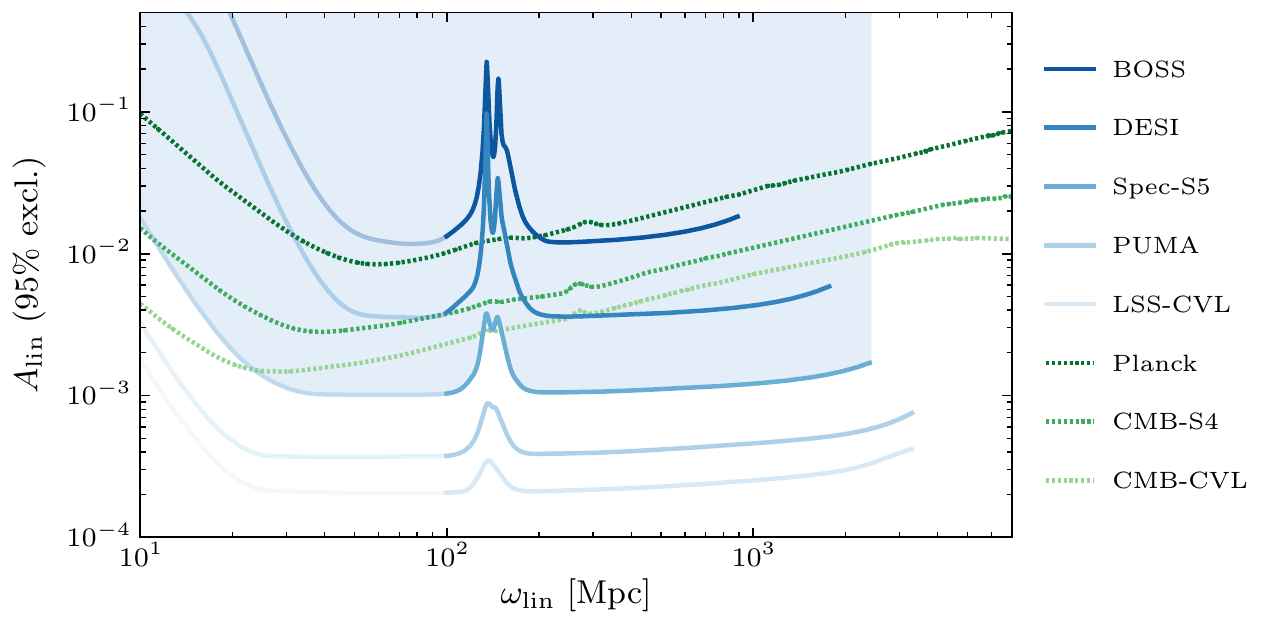}
    \caption{A Stage V Spectroscopic Facility would greatly increase our ability to detect primordial features in the matter power spectrum across a wide range of scales. This plot depicts the 95\% exclusion limit for the amplitude of linear features in the matter power spectrum, $A_{\rm lin}$, as a function of the scale at which that feature appears, $\omega_{\rm lin}$, for a variety of CMB (dashed) and LSS (solid) experiments.  The Spec-S5 curve (with corresponding detection area shaded) 
corresponds to MegaMapper using a maximum wavenumber $k_\mathrm{max} = 0.5h$Mpc$^{-1}$.  In addition, we show the constraints that would be obtained from cosmic-variance-limited (CVL) observations of LSS up to $z_\mathrm{max}=6$ over $\Omega = 20,000$deg$^2$ using a maximum wavenumber $k_\mathrm{max} = 0.75h$Mpc$^{-1}$ and of the CMB up to $l_\mathrm{max}^T =3000$ and $l_\mathrm{max}^P = 5000$ over 75\% of the sky and from a future 21cm line intensity mapping experiment, PUMA~\cite{PUMA:2019jwd}. Overall, LSS surveys have the potential to improve over the CMB sensitivity by more than an order of magnitude. 
    Figure adapted from \cite{2203.08128}.
    } 
\label{fig:alin}
\end{figure}

Primordial features in the power spectrum are a separate signal of physics beyond the standard models of cosmology and particle physics. These inflationary imprints are a manifestation of primordial dynamics that exhibits a significant departure from scale invariance and arise in broad classes of models, including both of inflation and its alternatives. Finding such inflationary signatures in cosmological observables would be a groundbreaking discovery that would open an entirely new window into the primordial universe. See a broader discussion in  \cite{2203.08128}.

Vanilla models of inflation predict almost Gaussian fluctuations with a nearly scale-invariant power spectrum. However, many models of the very early universe beyond the simplest incarnations of single-field slow-roll inflation generically predict departures from scale invariance. Since these deviations from the minimal power-law power spectrum of initial fluctuations are strongly scale-dependent, primordial features are typically oscillatory and/or localized in  momentum space. Observationally, primordial features could be imprinted in the spectrum of the cosmic microwave background, its anisotropies, all tracers of the large-scale structure of the Universe and the stochastic gravitational wave background.

Measurements of the matter power spectrum provide a probe of the initial power spectrum of fluctuations left behind after inflation, providing one of the few ways we can explore this phenomenon.  
The imprint of primordial features can be constrained (and potentially detected) with sufficiently-sensitive
measurements of the galaxy power spectrum, as recently demonstrated in \cite{1906.08758}, which inferred constraints slightly stronger than those from Planck CMB data.
Large-area surveys of the universe at high redshift are optimal for this work as they will measure clustering within an unprecedentedly large volume of the Universe, as discussed in \cite{2203.07506}.  
Fig~\ref{fig:alin} shows the ability of various surveys to constrain the presence of primordial features in the power spectrum.

The cosmological use of the large number of modes that are in principle accessible in galaxy surveys is
usually limited by gravitational non-linearities, baryonic physics on small scales and observational shot
noise. Recent advances in the theoretical understanding of these effects on the feature imprints in the LSS
spectra now allow to employ not only linear scales, but also those in the (weakly) nonlinear regime:
large-scale gravitational bulk flows can be resummed in perturbation theory and treated analytically, and
small-scale non-linearities should not impact these inflationary oscillations as long as their frequency is
high enough. This implies that primordial features with a high enough frequency can be searched for with the
high-redshift large volume 
of a future spectroscopic survey, increasing both the statistical sensitivity as well as systematic robustness of a potential discovery.
In addition, it suffices to model the oscillatory part of the power spectrum, i.e. we do not need to model its full shape, which is an easier problem and can be achieved to smaller scales than the full nonlinear treatment of biased tracers. Consequently, LSS constraints are independent of and competitive with those from the CMB anisotropies currently, and superior in future surveys.

\textbf{Multi-tracer techniques with high-density samples}

In the majority of cosmological measurements, the measurement is limited by the sample variance of the survey
volume. In fact, this is the main driver for the high redshift survey: to maximize the available volume and
thus the number of linear modes. However, for a certain subset of measurements, we can perform measurements by
using multiple samples over the same volume where the sample variance cancels for the right combination of
samples. This method has been proposed in late 2000s for non-Gaussianity \cite{0807.1770} and redshift-space
distortions \cite{0810.0323}. Similar methods in the context of combining weak lensing and CMB lensing were
studied a decade later \cite{1710.09465}. A renewed understanding of biasing of tracers have shown that these
early studies were likely overly optimistic, since there are white noise contributions from biasing that are
present at the largest scales. Continued progress on the theoretical front  is expected and these techniques
are considered very promising.

However, these methods only work if the galaxy shot noise is negligible, which implies very high number
densities. The current and upcoming generation of experiments do not have sufficient number densities to make
these studies realistic. However, the low-redshift sample of the Stage V experiment will be able to reach
interesting regime. In particular, this would potentially enable an independent probe of both primordial non-Gaussianity as well as structure growth.

\subsubsection{Exploring Important Problems}

A Stage V spectroscopic facility would also enable progress on a variety of important problems of current interest, using the same data that would be obtained to study cosmic acceleration.
 
\textbf{Constrain modified gravity models by constraining the slip parameter gamma}:
Modifications to GR have been proposed as an alternative explanation for the observed accelerated expansion of
the Universe at low-$z$. The phenomenology of such models can be very rich, but a common ingredient is that
the responses of the Newtonian gravitational potential $\Psi$ and spatial curvature Weyl potential $\Phi$ with
which massless particles interact gravitationally  may differ. These potentials are equal in magnitude in GR.
Redshift space distortions from spectroscopic surveys in combination with weak gravitational lensing (from
either CMB or photometric experiments) directly constrain the gravitational slip parameter $\gamma =
\Phi/\Psi$. The spectroscopic Stage V facility will reach $4-6\%$ errors on $\gamma$ at $2<z<4$ \cite{Sailer:2021}, and the high density low-$z$ sample will do better then $4\%$ at low-$z$. 
Weak lensing surveys traditionally use a different parameterization \cite{des-y3-3x2-ext}, $(\Sigma_0,\mu_0)$, where 
$\mu_0 \propto \Psi$ and $\Sigma_0 \propto (\Psi+\Phi)$. 
The spectroscopic Stage V facility measurements of $\gamma$ will be highly complementary to the Rubin/LSST measurements of ($\Sigma_0,\mu_0)$ and their combination will further enhance our understanding of the modified gravity landscape.


\textbf{Constraining the symmetry breaking pattern of inflation}: We return to measuring primordial
non-Gaussianities in the galaxy distribution, discussed earlier. Here we want to focus on other classes of PNG, and in particular equilateral PNG.
For single-field slow-roll inflation, the prediction is that the shape function peaks around the equilateral configuration with amplitude proportional to the scalar tilt which is beyond reach of near-future surveys. Fig~\ref{fig:fnl} shows the expected limits from Stage V like experiments. 
It is expected that theoretical developments will be able to lower those projected limits (cf. the recent
first measurements from BOSS data \cite{2201.07238,2201.11518}).

Equilateral PNG is a prime example of a very important observable for which a significant theoretical target is beyond the capability of a Stage-V spectroscopic survey, but for which the Stage V dataset will enable and propel rapid theoretical and analysis development. So far, there is no evidence for PNG for any shape or probe, with the tightest constraints being derived from Planck data. At the same time, there is a continued effort to mitigate and model astrophysical and nonlinear effects, spanning analytic, numerical and simulation-based approaches, to more efficiently extract primordial information from late-time observables. Together with the dramatic increases in observational sensitivity, future CMB and LSS analyses are projected to significantly improve the constraints on all PNG types, with potentially decisive implications for our understanding of inflation.

\textbf{Measure the sum of the neutrino masses:} Neutrino oscillation experiments constrain the difference in
the square of the mass of each member of a pair of neutrino mass eigenstates, $\Delta(m^2)$. However, this
leaves the actual mass of each eigenstate and their ordering ambiguous.
Since the nonzero masses of neutrinos are an indication of physics beyond the standard model, better
constraining these  masses and their hierarchy is an important problem in high energy physics. Refer to \cite{Abazajian:2022ofy}  for further discussion.

In contrast to neutrino oscillation experiments, cosmic spectroscopic surveys are sensitive to the \textit{total} mass of all varieties of neutrinos (i.e., the sum $m_1+m_2+m_3$).  This sensitivity arises because neutrinos affect the growth of cosmic density fluctuations in a scale-dependent manner.  
If the total mass in neutrinos is high, the power at small scales is reduced as neutrinos will not be
gravitationally bound to overdensities. 
However, neutrinos also affect the growth of clustering at large scales, as in the early universe they were highly relativistic and affected the growth of perturbations in the same way as radiation, but at later times they become non-relativistic and instead have a matter-like effect.  The strength of this effect again will depend on the total mass of all neutrino eigenstates.  
Since peculiar-velocity measurements directly measure the inflow of matter towards overdensities, they can provide direct measures of growth rates that complement inferences from the galaxy power spectrum \cite{1911.09121}. 

Both high-density/low-redshift and high-volume/high-redshift samples will enable improved constraints on neutrino masses.  At low redshift, we expect that being  simultaneously able to measure and compare the clustering of many different populations that each trace the matter differently to lead to an improved understanding of the relationship between the clustering of galaxies and of dark matter.  Such improvements could enable matter power spectrum information at small scales, where neutrino free-streaming should affect the observed signals, to be extracted. Similarly, the expected improvements to our understanding of the relationship between observed redshift-space distortions and the underlying flows of dark matter that should come from dense samples would enable neutrino mass constraints from redshift-space distortions to be interpreted with confidence.  At higher redshifts, the primary improvements to neutrino mass constraints will come from having better power spectrum measurements at large scales; high-$z$ spectroscopic samples from future surveys could cover an extremely large volume of the universe, improving power spectrum measurements at the largest scales.


\textbf{Tensions in measurements of the Hubble Constant:} Spectroscopic surveys have played a key role in
investigating tensions between measurements of the Hubble parameter $H_0$, which provides a measure of the
present-day cosmic expansion rate, based on low-redshift versus high-redshift measurements \cite{2203.06142}.
In particular, BAO measurements effectively determine distances using a scale calibrated using the cosmic
microwave background acoustic peak at high redshift, providing an ``inverse distance ladder'' calibrated in the early Universe.  Supernova distances can be calibrated to match the BAO scale at $z \sim 0.5$, allowing a CMB-based distance scale to be measured at redshifts as low as $z=0.02$ 
(as in \cite{1811.02376}).  Equivalently,  BAO-based determinations of $H_0$ can be compared to measurements made using standard candles calibrated with a low-redshift distance ladder. The eBOSS survey found that the resulting inverse-distance-ladder value of $H_0$ is consistent with the value inferred from Planck CMB measurements and inconsistent with local measurements
\cite{2007.08991}; the inverse-distance-ladder approach gives a value $H_0 = 67.87 \pm 0.86$.  If one is willing to assume a baseline \lcdm\ cosmology with no extensions, eBOSS obtains a value $H_0 = 67.35 \pm 0.97$ even without the incorporation of information from the CMB from the combination of BAO and Big Bang nucleosynthesis measurements.  Future surveys with better BAO measurements at higher redshifts should provide stronger constraints still from this CMB-independent approach.  
A Stage V spectroscopic facility could play an important role in exploring both the nature of and the physics underlying the tension in current Hubble parameter measurements. 

\textbf{Tensions in measurements of the amplitude of the matter power spectrum:} Experiments have repeatedly found that determinations of the amplitude of the matter power spectrum inferred from CMB temperature and polarization maps are in tension with measurements based on galaxy clustering and lensing, a subset of redshift-space distortions measurements, or the Sunyaev-Zel'dovich effect made at lower redshifts, as reviewed in 
\cite{2203.06142}.  Lower-redshift samples with Spec-S5 will include galaxies with a wide range of large-scale structure biases; however, since \textit{all these galaxies trace the same underlying web of matter}, the inferred matter power spectrum from different tracers at the same redshift should agree.  If not, we can infer the presence of systematics in current methods, and should give any apparent tension lower credence.  

The spectroscopic data set that would be collected by a Stage-V spectroscopic facility would enable many
techniques of measuring the amplitude of the matter power spectrum, not just using multiple tracers. 
Just inferring the power spectrum with observed clustering and a simple bias model is likely to be insufficient
without strong priors, but a full shape analysis is likely to succeed, and there is information in higher
order statistics such as the galaxy 3 and 4 point correlation functions.
Redshift-space distortions within the spectroscopic samples provide another constraint on the clustering of matter.

One can go even further by combining spectroscopic samples with measurements based on other datasets.  At lower redshifts, galaxy-galaxy lensing, which would combine spectroscopic samples in the foreground and Rubin Observatory (or space-based) lensing distortion measurements for background objects, will provide an another method of mapping the overall distribution of matter.  At higher redshifts ($z > 1$) the amplitude of CMB lensing around foreground spectroscopic objects will provide another means of mapping the distribution of matter. Cross-correlations with Sunyaev-Zel'dovich or X-ray maps can provide additional information.
By enabling high-precision measurements using a wide range of methods with disparate systematics, with
redshift coverage spanning from $z \sim 0.1$ to $z \sim 5$, surveys with a Stage V spectroscopic facility should play a key role in assessing the nature and redshift-dependence of tensions in measurements of the amplitude of the power spectrum.

\subsection{Enhancing LSST and CMB-S4 via Spectroscopy}

\label{sec:s5ss_enabling}

A Stage V spectroscopic facility can also enable improved constraints on cosmology by unlocking the full constraining power of near-future experiments such as the Vera C. Rubin Observatory and CMB-S4.  These new experiments will provide information-rich datasets, but will provide only limited information about redshift (in the case of Rubin Observatory) or none at all (CMB-S4).  Both new analyses of the large, wide-area surveys discussed above as well as smaller-area, more focused spectroscopic programs with new facilities can help to fill in the missing information and yield stronger constraints on cosmic acceleration from imaging-only projects at a fraction of their total cost. 

\textbf{Improving photometric redshifts from the Vera C. Rubin Observatory:}  As was described in \autoref{sec:photoz}, a set of at least 20-30,000 highly-secure redshift measurements for very faint galaxies will enable improved photometric redshift estimates for LSST, improving its cosmological constraining power significantly.  As can be seen in \autoref{table:photoz_times}, the proposed Stage V spectroscopic facilities would be extremely efficient at obtaining such a sample; \textbf{Spec-S5 has the potential to greatly increase the power of LSST with an investment of only a few months of observing time.} 

The proposed moderate-$z$, high-density samples that a Stage V spectroscopic facility would enable would provide multiple tracers of structure with different biases, allowing detailed reconstruction of redshift distributions via the cross-correlation methods described in \autoref{sec:photoz} with multiple cross-checks at the redshifts where the bulk of the LSST lensing signal will originate.  In turn, the proposed higher-redshift samples will improve the characterization of the higher-redshift tail of the LSST source distribution where DESI and 4MOST only have quasars and absorption systems.   As a result, a new Stage V spectroscopic facility would have the potential to retire one of the greatest potential sources of systematic errors in LSST cosmological analyses \cite{1809.01669}. 
Similarly, direct and cross-correlation measurements using these samples will probe intrinsic alignments at
higher redshifts, helping reduce the next leading source of systematic errors in LSST lensing analyses.

\textbf{Unlocking additional cosmological information via cross-correlations:} Measuring cross-correlation statistics (i.e., the correlation between one quantity and another as a function of separation or scale) that combine the large survey samples that new spectroscopic facilities would provide with measurements from Rubin Observatory or CMB experiments provides additional information that neither dataset can access on its own.  

We have already discussed above how measuring \textit{density} correlations between spectroscopic and photometric samples has the potential to provide detailed redshift information for Rubin Observatory studies.  However, by also measuring the correlations between the density of galaxies of given properties in a spectroscopic sample and the observed weak lensing shear from photometric objects in the background to those galaxies, we can also directly study the distribution of total mass around the spectroscopic objects, and hence infer the relation between these observed samples and the underlying dark matter.  In combination with the large-scale-structure bias information available in density cross-correlations, one can then study the underlying power spectrum of density fluctuations itself \cite{1207.1120}.  By providing large samples with a range of biases at $z<1.5$ and ample sample sizes up to $z=3+$, a future spectroscopic facility can enable multiple cross-checks of methods at lower redshift and directly infer the clustering of matter out to the highest redshifts probed by Rubin Observatory lensing.  

Cross-correlations between spectroscopic samples and CMB lensing maps should be powerful as well.  Whereas Rubin Observatory lensing analyses will measure shear using multiple photometrically-determined redshift bins, providing some redshift information on where the lensing mass is located, the CMB lensing signal all originates at very high redshift, with no redshift discrimination.  However, by measuring the cross-correlation between spectroscopic samples and the CMB lensing signal, it should be possible to reconstruct the matter power spectrum as a function of redshift, since the contribution of each redshift to the net signal can be measured directly. For such analyses, extending spectroscopic surveys to as high redshift as is feasible would be valuable, as the efficiency of CMB lensing is greatest at $z \sim 2$ but falls off slowly. One critical advantage of cross-correlation statistics for such analyses is that spectroscopic samples will have very different systematics from CMB lensing maps; such systematics will affect autocorrelations based on a single type of sample, but vanish when cross-correlations are measured \cite{1309.5388,1207.4543}.  

\subsection{Proposed Stage V Spectroscopic Facilities}

Several proposed dedicated facilities for Stage V-level spectroscopy were described in Snowmass Letters of Intent; we briefly summarize three of them here, MegaMapper, MSE, and SpecTel:

\textbf{MegaMapper:} The proposed MegaMapper facility would consist of a dedicated 6.5m diameter telescope coupled with 26,100 fiber positioners that take light to a set of DESI spectrographs covering 360-980 nm \cite{1907.11171,2209.04322}.  The telescope main mirror would be similar in design to those of the existing Magellan telescopes, but a hyperbolic secondary mirror and corrector lenses would enable a much larger field of view than is available at Magellan, totaling 7.1 square degrees.  MegaMapper was proposed by Lawrence Berkeley National Laboratory and the Carnegie Observatories as a dedicated future platform for cosmology surveys.  It is estimated that construction would take 7 years from project approval, making this the earliest feasible option for a massively multiplexed facility.  However, additional research and development is still needed to develop fiber positioners with the 6.5 mm center-to-center separation necessary to achieve the desired multiplex within the planned 1.2 m diameter focal plane (for comparison, DESI positioners have a 10.4 mm pitch).  We note that such development could benefit any of the proposed spectroscopic facilities.

\textbf{The Maunakea Spectroscopic Explorer:} The Maunakea Spectroscopic Explorer (MSE) would consist of a new 11.25-14m diameter telescope with a 1.5 square degree field of view, coupled to 4000-21,000 fibers \cite{1904.04907}.  One quarter of these fibers would be coupled to high-resolution spectrographs that are poorly suited for cosmic surveys, but the remainder would use moderate-resolution spectrographs covering at minimum 360-900 nm, with a subset feeding near-infrared spectrographs covering 1-1.8 $\mu m$.  Given the planned focal plane diameter of 1.3m, improvements to fiber positioner size beyond DESI would still be needed to achieve this multiplex.  MSE would replace the existing Canada-France-Hawai'i Telescope at the summit of Maunakea, utilizing much of the existing infrastructure, though a new and enlarged dome would be needed to accommodate recent designs.  MSE has passed a conceptual design review and the two-year preliminary design phase for it should begin in 2022; it is anticipated that science operations could begin within ten years if the schedule is technically-limited.

\textbf{SpecTel:} An European Southern Observatory (ESO)-sponsored study has proposed developing an 11.4m diameter spectroscopic telescope supporting a 5 square degree field of view and hosting 15,000 fibers.  Some fibers may be devoted to high-resolution spectrographs or integral field unit observing modes, but most would be coupled to moderate-resolution spectrographs with a wavelength range of 360 -- 1330 nm, suitable for the surveys discussed in this report.  The proposed design could achieve the nominal multiplex of 15,000 with DESI-sized fiber positioners, or higher numbers with a smaller pitch. This design delivers a worse image quality than MSE but enables a significantly larger field of view, advantageous for wide-area surveys.  Such tradeoffs as well as cost and schedule needs would be evaluated in the conceptual design process, which has not yet occurred for SpecTel.  

\begin{table}
\footnotesize
\newcolumntype{x}{>{\centering\arraybackslash\hspace{0pt}}p{1.2cm}}
\newcolumntype{y}{>{\centering\arraybackslash\hspace{0pt}}p{2.0cm}}
\newcolumntype{z}{>{\centering\arraybackslash\hspace{0pt}}p{2.7cm}}
\hspace*{-1.8cm}  \begin{tabular}{|p{2.2cm}|x|z|y|y|x|y|z|}
                  \hline &&&&&&&
                  \\
                  & Experiment type & Concept & Redshift Range & Primordial FoM & Time-scale & Technical Maturity & Comments \\
\hline
DESI & \scriptsize spectro & \scriptsize 5000 robotic fiber fed spectrograph on 4m Mayall telescope  & \scriptsize $0.1<z<2.0$ & 0.88 & now & operating & \\
\hline
Rubin LSST & \scriptsize photo &  \scriptsize \textit{ugrizy} wide FoV imaging on a 6.5m effective diameter dedicated telescope & \scriptsize $0<z<3$ & - & 2025-2035 & on schedule & \scriptsize Targeting survey for next generation spectroscopic instruments \\
\hline
SPHEREx & \scriptsize narrowband & \scriptsize Variable Linear Filter imaging on 0.25m aperture from space & \scriptsize $0<z<4$ & - & 2024 & on schedule & \scriptsize Focus on primordial non-Gaussianity \\
\hline
\hline
MSE+${}^\dagger$ & \scriptsize spectro & \scriptsize up to 16,000 robotic fiber fed spectrograph on 11.25\,m telescope &\scriptsize  $1.6 < z < 4$ (ELG+LBG samples) & $< 6.1$ & 2029- & high & \\
\hline
MegaMapper & \scriptsize spectro & \scriptsize 20,000 robotic fiber fed spectrograph on 6m Magellan clone & \scriptsize $2<z<5$ & 9.4 &2029- & high &  \scriptsize Builds upon existing hardware and know-how \\
\hline
SpecTel${}^\dagger$ & \scriptsize spectro & \scriptsize 20,000-60,000 robotic fiber fed spectrograph on a dedicated 10m+ class telescope & \scriptsize $1<z<6$ & $<23$ & 2035- & medium  &  \scriptsize Potentially very versatile next generation survey instruments\\
\hline
PUMA & \scriptsize 21\,cm & \scriptsize 5000-32000 dish array focused on intensity 21\,cm intensity mapping &  \scriptsize $0.3 <z<6$ & 85 / 26 (32K / 5K optimistic) & 2035- & to be demonstrated  & \scriptsize Very high effective number density, but $k_\parallel$ modes lost to foregrounds \\
\hline
mm-wave LIM concept & \scriptsize CO/[CII] LIM & \scriptsize 500-30000 on-chip spectrometers on existing 5-10m telescopes, 80-300\,GHz with R$\sim$300-1000 & \scriptsize $0<z<10$ & up to 170 &  2035 - & to be demonstrated & \scriptsize CMB heritage, can deploy on existing telescopes, very high effective number density, but $k_\parallel$ modes lost to foregrounds \& resolution \\
\hline
  \end{tabular}
  \caption{Table comparing current and next generation experiments capable of performing 3D mapping of the Universe. The upper part of the table shows existing and funded experiments, while the lower part is focused on proposed future facilities. See \cite{Sailer:2021} for further details. ${}^\dagger$ We have computed the FoM for MSE and SpecTel assuming they performed a full time LBG/LAE survey -- such a survey was not part of their proposals and those collaborations have not committed to doing any such survey.  For their proposed surveys the FoM is significantly lower. Adapted from \cite{2203.07506}. }
  \label{tab:experiments}
\end{table}

\subsection{Considerations for Evaluating Stage V Spectroscopic Facilities}
\label{sec:s5ss_facilities}

Any of the proposed Stage V spectroscopic facilities would represent a significant advance over what is possible with current resources and would enable progress on all of the science described in this section.  Given this, and that all three proposed facilities are still in the process of developing designs and collaboration models, it is not appropriate to select a specific implementation yet, but rather to establish a clear process and requirements for a final selection.

In general, the ability of these facilities to contribute to cosmic frontier science will be maximized if:

\begin{enumerate} 

\item The etendue of the system (i.e., the product of the collecting area and field-of-view, $A\Omega$) is as large as feasible while still maintaining good optical quality.  Increasing etendue will increase the speed of wide-area surveys, which are critical to the proposed science.

\item The focal plane area of the system is as large as possible (again, without sacrificing optical quality) in order to increase the number of fiber positioners that can be accommodated.  A minimum of 10,000 fiber positioners should be required to enable significant advances over what DESI can achieve, with 20,000 or more simultaneous positioners preferred. Fiber-densities of more than 10,000 per square degree are likely to be excessive for wide-area science cases, but if the instrument serves multiple science cases,  the number of targettable objects naturally increases, allowing higher fiber-density designs to be efficient.

\item The spectrographs used for cosmic acceleration surveys provide continuous coverage over the full optical window from 370 to 1000 nm, with wavelength coverage extending up to 1.6 $\mu$m in the infrared desirable but not absolutely required.  At wavelengths above 600nm spectral resolution should be sufficient to resolve the [OII] 3727 Angstrom doublet, providing secure redshift measurements from a single feature; this requires a resolution $R = \frac{\lambda}{\Delta \lambda} \sim 4000$ or above.

\item The collecting area of the facility should be at least as large as that of Rubin Observatory, in order to facilitate spectroscopy of faint targets (with larger collecting area preferable for faint-object science cases).  

\item All else being equal, a Southern hemisphere (or at minimum tropical) site is preferred in order to maximize synergies with the Rubin Observatory LSST and with CMB experiments.
\end{enumerate}

These considerations will need to be weighed against the amount of new funding needed for construction and operations in conjunction with other partners; the fraction of observing time that would be dedicated to surveys to study cosmic acceleration and dark matter; and the date when a facility would become available (e.g., LSST supernovae follow-up will not be feasible if LSST ends before construction of a facility is completed).  
Research and development on the miniaturization of fiber positioner systems would help to maximize the capabilities of a new facility when it is constructed by increasing multiplexing capabilities.

\section{Opportunity: Cosmological Physics with Extremely Large Telescopes (ELTs)}

\label{sec:elts}

The upcoming generation of extremely large telescopes (ELTs), including the European-led Extremely Large Telescope and the US-led Thirty Meter Telescope (TMT) and Giant Magellan Telescope (GMT) will be as transformative for astronomy as the new space based JWST. However, their designs provide fields of view that are small compared with the instruments capable of carrying out the wide field cosmological surveys of the last 25 years.  For the tasks they are designed for they are extremely capable machines, and they are the top priority in the recent Astro2020 decadal survey for astronomy and astrophysics. HEP would be remiss not to consider their use for cosmology.  Their excellent sensitivity and refined adaptive optics capabilities make them uniquely capable of following up faint targets discovered by dedicated survey telescopes. We outline some of the key opportunities identified by the community white papers below.

\textbf{Characterizing strong lens systems:} Strong lensing refers to distortions of the light from distant objects by the gravity of foreground objects that goes beyond low-order perturbations. Strongly lensed objects often exhibit multiple images; i.e., we see light from a single object appearing at several distinct points on the sky. The paths from source to the observer for these distinct points on the sky are of different lengths, so we see variability in one point occurring at the other points days to years later. The measured time delays between different light paths, when combined with a model f    or the lensing system, allows one to determine the value of the Hubble parameter. The time delay of a strongly-lensed     system is one of the very few dimensionful observables in cosmology.

Strongly-lensed quasar time delays have a long history of being studied and have been used to measure H$_0$ to $\sim 6\%$ \cite{2007.02941}.  Lens systems will require monitoring (requiring repeated imaging on smaller telescopes), but the modeling of lens systems also requires spectroscopic measurements of the redshift and velocity dispersions of the lens galaxy and extremely-high-precision measurements of the location of each image.  This latter information will be best obtained by adaptive-optics integral field unit spectroscopy on large telescopes and ELTs (depending on the brightness of the sources involved).  

The recent discovery of the first multiply lensed supernova SN \cite{1411.6009} has opened presented opportunities with a new class of objects.  Measuring time delays from lensed supernovae has many advantages over using quasars, 
Lensed supernovae require less monitoring and are less sensitive to microlensing \cite{1708.00003}, mass modeling systematics \cite{astro-ph/0211499}, and selection bias \cite{1605.08341}. Time delays from lensed supernovae also present opportunities to observe the earliest phases of supernova explosions, to infer cosmological parameters, and to map substructure in lens galaxies impacting dark matter physics, but many more systems are needed to achieve these goals.

\textbf{Photometric redshift training spectroscopy for LSST:} ELTs can also play a role in obtaining photometric redshift training/calibration spectroscopy for LSST.  Although their comparatively small field of view and lower multiplexing compared to DESI or PFS limits the sample size and area that can be surveyed at one time, their huge light-gathering power helps to make up for it.  As a result, ELTs could still play a role in photometric redshift training, as illustrated in \autoref{table:photoz_times}.  Based upon their expected instrument characteristics, TMT and E-ELT would require more than 400 dark nights to conduct the baseline LSST training survey, but GMT with the MANIFEST fiber positioner could achieve this in fewer than 200.  It is thus possible that ELTs could play a role in this work, potentially in concert with other facilities (e.g., obtaining infrared spectroscopy of objects that failed to yield redshifts in optical-only spectroscopy on smaller telescopes).  An intermediate option would be provided by the FOBOS spectrograph on Keck, which could perform the LSST baseline survey in $\sim 300$ dark nights, at substantially lower operating cost than an ELT.

\textbf{Characterizing Galaxy Clusters:} Clusters of galaxies are the most massive gravitationally bound structures in the universe. Their space density and clustering properties as a function of mass are extremely sensitive probes of cosmology, mostly via the growth of structure. To date the precision of the probe has been blunted by systematics in the determination of the cluster masses. Even the very good Sunyaev-Zeldovich selected samples (the SZ is a Thomson-scattering induced shadow on the CMB) require spectroscopic followup, and the optically selected samples from the DES require spectroscopic studies of contamination in the mass proxy and the use of extensive velocity dispersion measurements helps disentangle the effects of mass proxy covariance with weak lensing mass profile. The cluster cosmology measurement with the LSST data will be strengthened considerably with a dedicated campaigns aimed at observing a fair subsample of LSST galaxy clusters.
These studies are not well suited to the cosmological survey wide field massively-multiplexed spectrograph, as even the most ambitious of these don't provide a sufficient density of fibers on cluster to obtain the tens to hundreds of cluster galaxy spectra required without many visits. They are well suited to ELT spectroscopy, particularly using the GMT. A program to take 10 minute exposures on 10,000 clusters using 200 nights would allow detailed exploration of all of these systematics, a catalog of velocity dispersions, and caustic turn-around radius determinations, and more, perhaps in return for an upgraded multi-fiber spectroscopic instrument.

\section{Research and Development for Future Experiments}

\label{sec:r_and_d}

A balanced program should include operating experiments, projects that are being built, projects that are in the design stage (for which pathfinder projects may be appropriate), and long-term research and development to enable future projects. 

In Fig~\ref{fig:CF-summary} we show the scientific reach of many facilities that can provide significant constraints on the dark energy, light relics, and inflationary parameters that motivate much of our report. It no longer is possible to separate cleanly dark energy constraints into low redshift observations and inflation constraints into z=1000 observations. It is a feature of the 2021 Snowmass process that we realized how tightly connected the science aims of CF3, CF4, and CF5 are. In the figure one can see both CMB-S4 and Spec-S5 contributing constraints onto most of the parameters. CMB-S4 is clearly necessary to complete but is most fully described in the CF5 chapter. The next generation gravitational wave experiments are also important, as can be seen in the CF5, CF6 and CF7 chapters. Our community's interest in the GWO in this report is connected to using standard sirens to independently measure the H$_0$ and cosmology.  

We will present three pathfinder categories aiming at a Stage V spectroscopic facility, high precision measurements, and intensity mapping, that are ordered roughly chronologically for expected science payout. The first and the last are reflected in the Fig~\ref{fig:CF-summary},  but high precision is not represented on that projection of our science aims. All three represent our community's input.

\subsection{Pathfinders for a Stage V Spectroscopic Facility}

\textbf{DESI as a Bridge to the Next Generation}:
When it completes its five-year survey in 2026, the Mayall telescope coupled to DESI is expected to remain the most efficient instrumentation available for wide-area spectroscopic surveys until such time as a Stage V spectroscopic facility is built.  Given the modest etendue and multiplex of Mayall/DESI compared to what next-generation telescopes should provide, it cannot match the surveys that they would make possible.  Nevertheless, DESI should play several important roles in the intervening period.  These should include:

\label{sec:desi2}

\begin{itemize}

\item Performing prototype surveys that develop new target classes and observing modes for future facilities, while simultaneously producing new cosmological constraints.  This would follow the successful model of the recent eBOSS survey, which applied prototype selection methods that were being explored for DESI using the leading spectroscopic survey capability available, specifically the SDSS telescope with the BOSS spectrographs.  The eBOSS survey enabled these new selection techniques to be explored and assessed, and delivered improved constraints on cosmic acceleration models.   Importantly, it also helped to support the ongoing advancement of new analysis methods which are now being applied to DESI, as well as the development of a cadre of junior scientists who were well-prepared to contribute to the next generation of survey spectroscopy.

\item Targeted pursuit of the most efficient science opportunities.  Given its multiplexing and etendue, it is not feasible for DESI to simultaneously undertake multiple, dense surveys as has been proposed for Stage V facilities.  However, it would be feasible to pursue a selected, high-value subset of that science in order to advance our understanding of cosmic acceleration sooner; some possibilities are discussed in white papers submitted to this topical group \cite{2203.07291,2203.07506}.  As an example, it would be feasible to conduct a survey of $z>2$ objects (including both Lyman Break Galaxies and Lyman Alpha Emitters) over 3000 square degrees with the DESI instrument, if sufficient imaging for target selection is available.  Such a survey would cover a comparable amount of volume to the DESI galaxy samples, but at higher redshifts.  If a Stage V facility is not going to be available in the early 2030s, there will be correspondingly more need for DESI to contribute in this way. 

\item Maximizing the science from near-future imaging surveys.  As discussed in \autoref{sec:s5ss_enabling}, massively-multiplexed spectroscopy can enhance the science output from planned imaging surveys such as the Rubin Observatory LSST and CMB-S4.  In many cases these needs are time-sensitive -- particularly in the case of transient spectroscopy as for type Ia supernovae, but the quality of LSST photometric redshifts will also be limited by the availability of training spectroscopy.  Because of this, the complementary of DESI observations to planned imaging experiments and the time urgency of those observations should be given substantial weight in developing future DESI programs.  In some situations, DESI may improve cosmological constraints most per unit time via spectroscopic surveys themselves, but in others DESI data may deliver a higher impact by greatly improving the constraining power of other datasets; it is likely that a combination of the two strategies will be optimal.  
\end{itemize}

\textbf{Fiber Positioners:}
A Stage V spectroscopic facility features a large aperture optical/IR telescope with a focal plane designed for massively multiplexed spectroscopy.
The multiplexing is limited by the total physical focal plane area and the size of each robotic fiber positioner; the smaller the positioners are, the more of them can fit within a given-size focal plane.  
A spectroscopic Stage V facility would have a focal plan with $\gtrsim 20,000 - 50,000$ fibers. 

The newest generation of high-multiplex spectrographs, including both DESI and the Subaru Prime Focus Spectrograph (PFS), use a set of robotically controlled motors to move optical fibers to the positions of the objects.
The center-to-center distance between positioners is referred to as pitch.
DESI has a fiber pitch of 10.4mm, and Subaru/PFS has an 8mm pitch.  For these designs the pitch is primarily determined by the size of the robotic motors used to position fibers.  Smaller pitch positioners make possible more spectroscopic fibers for the same focal plane size. Thus development of smaller pitch positioners are a high priority. Likewise, the DESI experience suggests that the improved reliability of these positioners over year time scales is a priority.

If smaller positioners are developed, a focal plane making use of them could potentially be deployed at Mayall Observatory (coupled to the current DESI spectrographs combined with new ones to accommodate the greater total number of fibers) first, and then moved to Spec-S5 later, if the focal plane sizes and input beams are compatible.  This would enable some progress on Spec-S5 science to be made before a new telescope is completed, taking advantage of the rapid development and deployment cycle time from DESI and the experience gained from that project.  The Snowmass white paper \cite{2209.03585} discusses such scenarios in more detail.

\textbf{Skipper CCDs:}
Skipper CCDs are charge-coupled devices that have an output readout stage that allows multiple non-destructive measurements of the charge packet in each pixel of the array thanks to a floating gate output sense node. 
This allows Skipper CCDs to achieve sub-electron readout noise either over the entire detector area or over specific sub-regions of the detector that may be pre-defined before the exposure or dynamically configured based on the first charge measurement in each pixel.
Skipper CCDs are at a fairly mature stage of technical development and have been used in a wide range of  particle physics experiments.  Using Skipper CCDs can be beneficial for spectroscopic measurements in multiple ways.

First, the signal-to-noise ratio of spectroscopic observations of faint astronomical sources at wavelengths where the background noise from the sky is low can be limited by detector readout noise. In particular, current medium- to high-resolution spectroscopy at shorter optical wavelengths is in this regime; as a result, significant gains could be achieved through reductions in readout noise to $< 1$ e$^-$ rms/pix, which Skipper CCDs would make possible. For samples of objects where the crucial  measurements occur at blue wavelengths (e.g., Ly-$\alpha$ forest lines in high-$z$ quasars), reducing the readout noise to $0.5$ e$^-$ rms/pix would result
in a $\sim20\%$ increase in survey speed -- i.e., what would currently be a five-year survey would achieve its goals in four years. 

Furthermore, multi-object spectroscopic facilities generally will observe a set of objects of widely varying brightness simultaneously, with the same fixed exposure time for all targets. 
As a result, the ability to control readout noise dynamically that Skipper CCDs provide would provide photon counting when needed for faint sources, but will not waste time with repeated readouts on bright sources that are shot-noise dominated.

The major hurdle when applying Skipper CCDs to future spectroscopic instruments comes from the increased readout time needed to perform multiple independent samples.  Several avenues toward reducing the readout time of Skipper CCDs are being pursued at Fermilab and LBNL.  Further work in this area, including pathfinder projects, would enable them to be deployed in the Spec-S5 spectrographs.

\textbf{Germanium CCDs:}
Silicon CCDs are at this point a mature technology for measurements in the optical regime spanning $3,600 < \lambda < 10,000$\,\AA, but the effective band-gap around 1 eV limits their effectiveness at redder wavelengths. The primary
spectroscopic feature used to determine redshift in higher-$z$ galaxy surveys is a doublet of lines associated with forbidden transitions in
singly-ionized oxygen ([OII]); because it is a doublet, identifying this feature alone is sufficient for a secure redshift measurement. These [OII] emission lines occur at 3,727\,\AA\ in the galaxy restframe,
causing the signal to appear beyond the 10,000\,\AA\ cutoff in a silicon detector for galaxies at redshifts $z > 1.6$. 
Germanium CCDs can be processed with the same tools used to build silicon imaging devices, show promise for read noise and sensitivity comparable to that of silicon detectors, and offer a high quantum efficiency to wavelengths as red as 1.4 microns when cooled to 77 K. This increase in wavelength coverage would allow spectroscopic identification of [OII] emission
lines to $z = 2.6$, enabling a factor of two increase in volume over what is accessible in the DESI galaxy sample.

However, fabrication of germanium CCDs faces several challenges that need to be addressed before such devices
can be integrated onto large focal planes. 
Several processes in doping, etching, and film deposition are similar to those in silicon CCD fabrication, and may be compatible with current fabrication facilities. 
However, water solubility and low-temperature limitations result in the need for changes in gate-electrode technologies. 
In addition, there is only one wafer vendor in germanium and further investigation is required to ensure that purity requirements can be met at scale needed for production on large wafers. 
Finally, germanium is higher density than silicon and requires a full assessment of handling and packaging techniques. Pathfinder projects to develop germanium devices could significantly increase the capabilities of a Spec-S5 via spectrograph upgrades if they are not technologically mature when the facility is first constructed.

\subsection{Pathfinders for High-Precision Measurements} 

\textbf{High Precision Spectroscopy:}
Spectroscopy with high wavelength precision has many uses in astronomy; it has enabled many exoplanet discoveries and more recently has been used to explore dark matter dynamics in the Milky Way Halo as traced by the velocities and accelerations of stars. If the precision of radial velocity measurements could be further increased, it should be possible measuring the expansion of the universe \emph{directly} by observing the average redshift of objects increasing with time, a phenomenon commonly referred to as redshift drift. The size of this effect is redshift-dependent but is of order $\Delta \lambda / \lambda \sim 10^{-10}$ per year. 

\begin{figure}       
\begin{center}
\includegraphics[scale=0.6]{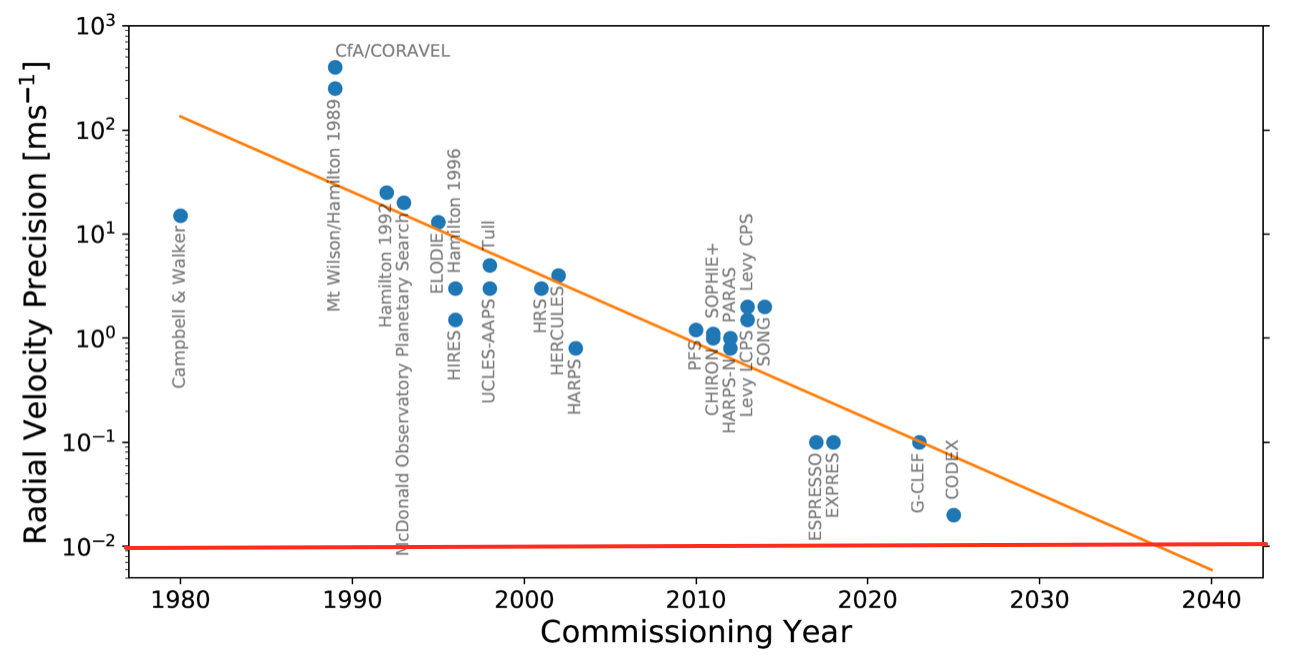}
\caption{The precision of radial velocity measurements has improved rapidly over the past three decades.  Plotted is the radial velocity precision achieved or expected for various recent or upcoming instruments, as a function of the year in which each instrument was commissioned (adapted from Silverwood \& Easther 2019 and taken from \cite{2203.05924}). Detecting cosmological redshift drift would require measurements with $\sim 1$ cm s$^{-1}$ precision (red line) with stability over timespans of years to decades; proposed pathfinder projects will soon approach that limit.}
\label{fig:RVprecision}
\end{center}
\end{figure}

The current state of the art in precision spectroscopy is approaching the levels of precision and instrument stability required to enable such measurements.  One promising technology being explored at LLNL is externally dispersed interferometric (EDI) spectroscopy \cite{2203.05924}. Redshift drift measurements would offer a very clean, direct measurement of the Hubble parameter, potentially enabling a resolution of current tensions in its measurement \cite{1907.05151}. Longer term, drift measurements would also enable a unique probe of the strength of cosmic acceleration. Figure \ref{fig:RVprecision} illustrates that, based on the historical rate of radial velocity precision improvements and the capabilities of currently planned instruments, we can anticipate that the precision needed to detect cosmological drift should be reached over the next two decades.  Efforts to deploy pathfinder projects over the next decade would help to enable this goal to be reached.

\textbf{High Precision Astrometry:}
High precision astrometry (i.e., precision measurements of apparent positions on the sky) also has the potential to enable potentially transformative new cosmological probes.  For example, distance quasars should exhibit apparent movements perpendicular to our line of sight due to cosmological parallax.  This effect can be further boosted for strongly lensed quasars as the precise line of sight changes over time \cite{2203.05924}.  At similar order of magnitude, distant objects should also exhibit true astrometric movements towards over-densities that could in principle be observed through cross-correlations of density tracers. In addition to these impacts relevant for studies of cosmic acceleration,  precision astrometric measurements in conjunction with precision radial velocities from the advances described above could enable a full 3D mapping of the motions of stars in the Milky Way, providing a more complete probe of the distribution of dark matter within our Galaxy.

Astrometric motions of sources at cosmological distances could be detected by planned next-generation facilities at radio wavelengths (such as ngVLA) as well as in the optical using the upcoming generation of extremely large telescopes. On the instrumentation front, new dedicated techniques are starting to be explored, employing quantum-assisted measurement techniques (e.g., optical interferometry using entangled photons). These methods have strong synergies with existing DOE investments into quantum technologies.

\subsection{Intensity mapping}
Intensity mapping represents an orthogonal approach to cosmological surveys, relying on instruments that give us no ability to isolate individual objects as in traditional galaxy surveys, but instead measure variations in aggregate intensity due to the combined emission from many objects, averaged over large scales. This aggregate intensity varies across the sky and with redshift, tracing the fluctuations in the underlying dark matter density that modulates the observed number density of (and hence aggregate emission from) galaxies.

Intensity mapping methods have primarily been employed for surveys at low frequencies where the limiting factor is the resolution of the instrument, but they can also be applied to work with objects that are below the detection threshold of a given survey. There are two currently promising avenues for pursuing such strategies:

\textbf{21\,cm intensity mapping:}
The 21\,cm spin-flip transition of neutral atomic hydrogen provides an ideal target for intensity mapping. Neutral hydrogen is present in most galaxies and can be observed using appropriately optimized radio instruments. The advances in commercial radio-frequency instrumentation driven by the needs of the wireless industry has enabled extremely cost-effective interferometric radio telescopes that could revolutionize survey cosmology. Modulations of 21\,cm intensity on large scales effect has been detected in cross-correlation by CHIME \cite{2202.01242} and MeerKAT \cite{2206.01579}.  

DOE laboratories have developed a concept for a future intensity mapping project, the Packed Ultra-wideband Mapping Array (PUMA) \cite{PUMA:2019jwd, 2002.05072}. A white paper submitted to this topical group \cite{2203.07506} has shown that the nominal performance of PUMA would enable unprecedented accuracy in measuring the expansion history and exploring inflationary signatures at higher redshifts than Spec-S5 can reach, as illustrated in e.g., Fig~\ref{fig:ode}.

However, 21\,cm intensity mapping is far from being ready for major investment; there is a large gap between the potential science yield from PUMA and the actual reach of the current generation of experiments. This gap can be traversed by pathfinder programs that enable the development of technologies and methods for the future.  This funding can be applied along two independent and largely orthogonal axes as outlined in PUMA Snowmass letter of interest (see also \cite{2002.05072}). First, as a part of long term strategy,  it is important to maximize the scientific return on investment from the current generation of precursor experiments, many of which suffer from under-staffing of the science effort. A strategic investment by DOE to enter into collaboration with one of the current generation experiments with the specific goal of contributing to the simulation and analysis effort would be a well suited to DOE's institutional strengths. Additionally, intensity mapping experiments would benefit from development of low-cost and low-energy-consumption implementations of the hardware that forms the generic building blocks of RF infrastructure: specifically, digital channelizers, correlators and spectrometers. This work could have significant overlap with instrumentation development for light sources and particle accelerators.

The modern universe 21\,cm is highly synergistic with the early universe radio cosmology efforts (nominally covered by the CF5 topical working group). LuSEE-Night is a funded pathfinder experiment that will be delivered to the lunar far side by the NASA Commercial Lunar Payload Services (CLPS) in 2025 and perform pioneering measurements of the radio sky at frequencies below 50MHz. The commonalities extend to both the development of hardware (for example low-power digital processing elements are required on Earth due to large number of elements and on the far side of the Moon due to general scarcity of power) as well as development of data analysis and calibration methods.

\textbf{Millimeter-wave intensity mapping:}
Rest-frame far-IR lines such as the CO rotational transitions and the [C{\sc ii}] fine-structure line can also be used for intensity mapping at millimeter wavelengths, from 10--600~GHz \cite{Kovetz17, 2203.07258}. About half of the optical light in the universe has been absorbed by dust in galaxies and re-emitted in the far-IR; these lines are therefore relatively bright due to the high abundance of carbon and of CO in cold molecular clouds, and have been detected in individual galaxies out to $z\sim 7$. Pathfinder experiments have detected CO in intensity mapping-style observations at $z\sim 3$ \cite{2008.08087}. 

The frequency range in which these lines may be detected coincides with the bands where ground-based CMB experiments operate---for example, a single receiver operating from 80--300~GHz would be sensitive to a combination of CO and [C{\sc ii}] lines spanning $0 < z < 10$. Because of this, mm-wave LIM experiments have the potential to reach mode counts beyond Stage V surveys and complement the science reach of PUMA. The measurement technique is similar to CMB experiments, requiring low-noise mapping of faint, diffuse emission over large sky areas with low systematics, but instead of broadband bolometers they would use spectrometers. Analog on-chip spectrometers have been under development for some time and pathfinder experiments are now being deployed to demonstrate their use \cite{2111.04631}. A primary advantage of performing intensity mapping at millimeter wavelengths is that it can profitably re-use existing CMB infrastructure, including existing telescopes and observatories. Detector development similarly takes advantage of several decades' experience in mass-producing photon noise limited, high-density mm-wave cryogenic detectors (which is now being scaled up to the extremely large detector counts needed for CMB-S4).  For this effort to become viable, targeted investment in developing on-chip spectrometers with competitive focal plane density, frequency resolution, and multiplexing is needed, along with on-sky demonstrations and verification of analysis techniques using staged pathfinder deployments. 


\section{Conclusion}

As we have described, projects at a wide variety of scales can help us to make progress on both investigating
the nature of cosmic acceleration in the early and late universe  and investigating other cosmological phenomena simultaneously.  Given the
richness of the datasets that they would provide and the many unique cosmological probes that they would
enable, a Stage V spectroscopic facility represents the greatest opportunity to advance our understanding;
such a facility would also provide new probes of the nature of dark matter, as described in Chapter 3 of this
report.  However, there exist other smaller-scale opportunities to enhance our understanding that can be
undertaken in the near term, including smaller-scale spectroscopic surveys with the DESI instrument and
efforts to obtain complementary data to enhance the cosmological constraining power of LSST.  These short-term
projects should be augmented by active development of pathfinders to explore new techniques and instrumentation concepts that could enable novel cosmological experiments that would begin operation in the later 2030s or beyond.  The Vera C. Rubin Observatory similarly could enable new opportunities to make progress once LSST is completed, though the time is not yet ripe to evaluate the possibilities.

A balanced portfolio of efforts should simultaneously enable us to enhance our understanding of cosmic
acceleration, including both tests of dark energy models and modifications to GR; probe the origins of cosmic
inflation by testing for non-Gaussianity and features in the primordial spectra; 
place stringent constraints on the radiation density and potential light relics beyond the standard model; 
provide strong constraints on the sum of neutrino masses; test for the presence of dark energy-like components in early phases of the Universe's history; and help to resolve current tensions in measurements of the Hubble parameter and the amplitude of matter density fluctuations.  The modern universe is rich in cosmological information; it will be up to us to take advantage of the opportunities it provides to improve our understanding of fundamental physics.

\section*{Acknowledgements}

The conveners of the CF04 topical group wish to thank all of the fellow physicists who have contributed to the group's discussions as well as to the letters of intent and white papers that underly the report's findings.  We particularly wish to thank those who reviewed drafts of this report and made suggestions that have helped to improve it, including Julien Guy, Michael Levi, Eric Linder, Saul Perlmutter, and David Schlegel. 







\bibliographystyle{Cosmic/CF04/JHEP}
\bibliography{Cosmic/CF04/cf04ref}

\end{document}